\documentclass[longauth]{aa} 

\usepackage{graphicx}
\usepackage{natbib}
\usepackage[varg]{txfonts}
\usepackage{multirow}
\usepackage{longtable}
\usepackage{xcolor}
\usepackage{booktabs}
\bibpunct{(}{)}{;}{a}{}{,} 
\usepackage{hyperref}
\begin{document}

   \title{Varying linear polarisation in the dust-free GRB\,210610B}

   \author{J. F. Ag\"u\'i Fern\'andez
          \inst{1}
          \and
          A. de Ugarte Postigo \inst{2,3}
          \and
          C. C. Th\"one \inst{4}
          \and
          S. Kobayashi \inst{5}
          \and
          A. Rossi \inst{6}
          \and
          K. Toma \inst{7, 8}          
          \and
          M. Jel\' inek\inst{4} 
          \and
          D.~A.~Kann \inst{1, 9}
          \and
          S. Covino \inst{10}
          \and
          K. Wiersema \inst{11}
          \and
          D. Hartmann \inst{12}
          \and
          P. Jakobsson \inst{13}
          \and
          A. Martin-Carrillo \inst{14}
          \and
          A. Melandri \inst{15, 10}
          \and
          M.~De~Pasquale \inst{16}
          \and
          G. Pugliese \inst{17}
          \and
          S. Savaglio \inst{6, 18, 19}
          \and 
          R. L. C. Starling \inst{20}
          \and
          J. \v{S}trobl \inst{4}
          \and
          M. Della Valle \inst{21}
          \and
          S. de Wet \inst{22}
          \and
          T.~Zafar \inst{23, 24}
          }

   \institute{Instituto de Astrof\'isica de Andaluc\'ia, Glorieta de la Astronom\'ia s/n, 18008 Granada, Spain\\
              \email{feli@iaa.es}
        \and
            Observatoire de la C\^ote d'Azur, Universit\'e C\^ote d'Azur, Boulevard de l'Observatoire, 06304 Nice, France
        \and
            Aix Marseille Univ, CNRS, CNES, LAM Marseille, France
        \and
            Astronomical Institute of the Czech Academy of Sciences (ASU-CAS), Fri\v cova 298, 251 65 Ond\v rejov, Czechia
        \and
            Astrophysics Research Institute, Liverpool John Moores University, 146 Brownlow Hill, Liverpool, L3 5RF, United Kingdom
        \and
            INAF – Osservatorio di Astrofisica e Scienza dello Spazio, Via Piero Gobetti 93/3, 40129 Bologna, Italy    
        \and
            Frontier Research Institute for Interdisciplinary Sciences, Tohoku University, Sendai 980-8578, Japan
        \and
            Astronomical Institute, Graduate School of Science, Tohoku University, Sendai 980-8578, Japan 
        \and
            Hessian Research Cluster ELEMENTS, Giersch Science Center, Max-von-Laue-Strasse 12, Goethe University Frankfurt, Campus Riedberg, 60438 Frankfurt am Main, Germany
        \and
            INAF - Brera Astronomical Observatory, Merate, Italy
        \and
            Centre for Astrophysics Research, University of Hertfordshire, Hatfield, AL10 9AB, United Kingdom
        \and
            Clemson University, Department of Physics and Astronomy, Clemson, SC 29634, United States of America
        \and   
            Centre for Astrophysics and Cosmology, Science Institute, University of Iceland, Dunhagi 5, 107 Reykjav\' ik, Iceland
        \and
            School of Physics and Centre for Space Research, O'Brien Centre for Science North, University College Dublin, Belfield, Dublin 4, Ireland
        \and        
            INAF - Rome Astronomical Observatory, via Frascati 33, I-00078, Monte Porzio Catone, Italy
        \and
            University of Messina, MIFT Department, Via F. S. D'Alcontres 31, 98166 Messina Italy
        \and
            Astronomical Institute Anton Pannekoek, University of Amsterdam, PO Box 94249, NL-1090 GE Amsterdam, The Netherlands
        \and
            Physics Department, University of Calabria, Via P. Bucci, 87036 Arcavacata di Rende, CS, Italy
        \and
            INFN-Laboratori Nazionali di Frascati, Via Enrico Fermi 54, 00044 Frascati, RM, Italy
        \and
            School of Physics and Astronomy, University of Leicester, University Road, Leicester, LE1 7RH, United Kingdom 
        \and
            Capodimonte Astronomical Observatory, INAF-Napoli, Salita Moiariello 16, 80131 Napoli, Italy
        \and
            Inter-University Institute for Data Intensive Astronomy \& Department of Astronomy, University of Cape Town, Private Bag X3, Rondebosch 7701, South Africa           
        \and
            School of Mathematical and Physical Sciences, Macquarie University, NSW 2109, Australia 
        \and
            ARC Centre of Excellence for All Sky Astrophysics in 3 Dimensions (ASTRO-3D), Australia
             }

   \date{Received XXXXXX XX, XXXX; accepted XXXXXX XX, XXXX}
 
  \abstract
   {Long gamma ray bursts (GRBs) are produced by the collapse of some very massive stars, which emit ultra-relativistic jets. When the jets collide with the interstellar medium they decelerate and generate the so-called afterglow emission, which has been observed to be polarised.}
   {In this work we study the polarimetric evolution of GRB\,210610B afterglow, at $z=1.1341$. This allows to evaluate the role of geometric and/or magnetic mechanisms in the GRB afterglow polarisation.}
   {We observed GRB\,210610B using imaging polarimetry with CAFOS on the 2.2 m Calar Alto Telescope and FORS2 on the $4\times8.1$ m Very Large Telescope. Complementary optical spectroscopy was obtained with OSIRIS on the 10.4 m Gran Telescopio Canarias. We study the GRB light-curve from X-rays to optical bands and the Spectral Energy Distribution (SED). This allows us to strongly constrain the line-of-sight extinction. Finally, we study the GRB host galaxy using optical/NIR data to fit the SED and derive its integrated properties.}
   {GRB\,210610B had a bright afterglow with a negligible line-of-sight extinction. Polarimetry was obtained at three epochs: during an early plateau phase, at the time when the light curve breaks, and after the light curve steepened. We observe an initial polarisation of $\sim4$\% that goes to zero at the time of the break, and then increases again to $\sim2$\% with a change of the position angle of $54\pm9$ deg. The spectrum show features with very low equivalent widths, indicating a small amount of material in the line-of-sight within the host.}
   {The lack of dust and the low amount of material on the line-of-sight to GRB\,210610B allow us to study the intrinsic polarisation of the GRB optical afterglow. We find the GRB polarisation signals are consistent with ordered magnetic fields in refreshed shock or/and hydrodynamics-scale turbulent fields in the forward shock.}

   \keywords{Gamma-ray Bursts: individual: GRB\,210610B --
                Polarisation --
                 Techniques: polarimetric
               }

   \maketitle
%

\section{Introduction}

Gamma-ray bursts (GRBs) are among the most energetic electromagnetic explosions that have been observed in the Universe. These events have two main emission episodes, the prompt- and the afterglow-emission phases. The prompt emission represents the first electromagnetic emission observable and is dominated by gamma-ray photons lasting seconds to minutes after the burst onset. The afterglow has a synchrotron spectrum ranging from radio to gamma-rays and evolves in time during much longer time spans.

GRBs are typically classified as short or long according to their measured T$_{90}$\footnote{The T$_{90}$ is the time-span in which a GRB releases between 5\% and 95\% of the total energy during the prompt phase.} duration in gamma-rays \citep{Kouveliotou1993}. Short GRBs (sGRB) are associated with the coalescence of two compact objects, typically have a T$_{90}$ duration lower than 2 seconds and a hard X-ray spectrum. The discovery of a sGRB associated with the gravitational wave (GW) detection GW\,170817 definitively linked a sGRB with the merger of two neutron stars \citep[][]{Abbott2017}. 
On the other hand, Long GRBs (lGRB) are those that show T$_{90}$ longer than 2 seconds and softer spectra on the prompt emission phase. These are cataclysmic events associated with the collapse of massive stars and are associated with the detection of Broad-Line (BL) Type Ic Supernovae (SNe) \citep[see e.g.][]{Galama1998, Hjorth2003}. Lately, several events detecting a KN emission in a burst with several tens of seconds duration have cast doubts on the 2 seconds division as the unique criteria to distinguish between sGRBs and lGRBs \citep{Rastinejad2022, Troja2022, Yang2022, Gompertz2022,Levan2023a,Levan2023b}.

In a GRB, the prompt emission is powered by a newly formed compact object fed by the surrounding material. Accretion onto this compact object launches ultra-relativistic jets, in which the prompt emission is generated through internal dissipation processes including internal shocks \citep[see e.g.][]{Rees1994, Sari1997, Kobayashi1997}. This prompt high-energy emission releases isotropic-equivalent energies that can reach up to 10$^{55} \textnormal{erg}$ \citep[see e.g. ][]{Burns2023}, although the real energy released can be several orders of magnitude smaller due to the jet collimation.

Polarimetry is an essential tool to explain GRB physics. Using this technique, we can test models that include magnetic fields and geometrical characteristics that are at play and how they evolve throughout the different phases of the GRB. The study of the prompt emission polarisation and its temporal evolution can help to understand how the jet is powered by the central engine. The long and extremely bright GRB\,221009A was observed during the prompt and afterglow emission phase by the Imaging X-ray Polarimetry Explorer (IXPE) and only upper limits were obtained \citep{Negro2023}. Until now, the studies on the prompt emission polarisation with instrument calibrated to perform polarimetry are scarce. The Astrosat CZTI mission shows the prompt emission as highly polarised while for POLAR, GRBs are lowly or non-polarised \citep[see e.g. ][]{Gill2021}.

The material ejected through the jets eventually collides with the circumburst material, decelerating as interacts with it. These forward shocks generate a broadband synchrotron emission, known as the afterglow. This bright shock can be observed during days or even months in the case of radio frequencies. In certain cases, a reverse shock that propagates backwards within the relativistic jet can be observed at early times \citep[see e.g. ][]{Meszaros1997,Piran1999}. Different models predict this emission to be polarised. \citet{GruzinovWaxman1999} proposed a model in which the afterglow would show patches with locally ordered magnetic fields randomly oriented but with many of them sharing a common direction leading to a global low level of polarisation of the GRB afterglow.

The number of studies using optical linear imaging polarimetry and spectropolarimetry techniques is still very small \citep[see e.g. ][]{covinogotz16}. The first GRB with a polarisation detection on the afterglow was GRB\,990510, with a polarisation degree (PD) of 1.7\,\%, 0.7 days after detection \citep{Covino1999, Wijers1999}. Many efforts have been carried out to increase the sample of polarimetric measurements, however, the faintness of these objects, their fast evolution, and the large observing times required for a good signal-to-noise ratio leave us with less than 20 GRBs with a measurement above 3-$\sigma$. Fast reaction is crucial to observe the first GRB phases where one could expect a higher PD. The highest measured value was for GRB\,120308A with a PD of 28\,\% 5 minutes after the burst, which rapidly decreased to 16\,\% as the afterglow evolved \citep{Mundell2013}. A high PD of $\sim$10\,\% was also measured for GRB\,020405, GRB\,090102 and GRB\,091208B \citep{Bersier2003, Steele2009, Uehara2012}. All of them were observed earlier than 0.01 days after the GRB detection. Observations at later stages shows this sources as lowly-polarised with some degree of variability throughout their light-curve evolution. The most detailed example is GRB\,030329 \citep{greiner2003} with a PD evolving somewhat randomly from 0.91\,\% at 0.5 days to 1.4\,\% 37.5 days after burst.

Afterglow polarisation must be understood together with the surrounding environment. Dust in the line-of-sight can change the observed PD and prevent the measurement of the intrinsic polarisation \citep{Lazzati2003}. Also, this measurement by itself lacks completeness if it is not followed by the study of the light-curve evolution. To understand whether there is or not a light curve break and how the polarisation behaves before, during and after the break is crucial to distinguish between the proposed models to explain GRB jet physics \citep[see e.g.][]{covinogotz16}.

In this paper we present a comprehensive study of GRB\,210610B, its afterglow emission and its environment using different techniques. It was a bright long GRB for which polarimetric, spectral and photometric observations were secured. We also observed the putative host galaxy in order to characterise it and put into the GRB context. This work is structured as follows: In section \ref{sec:observations} we present the observations of both, the afterglow and the underlying galaxy. In section \ref{sec:results} we present the results of the analysis of the afterglow linear polarimetry, the light-curve and its spectrum, as well as the analysis of the galaxy. In section \ref{sec:discussion} we discuss the results setting the framework in which GRB\,210610B is embedded. We also put the results into context of GRB afterglow polarimetry measurements. Finally, in section \ref{sec:conclusions} we present the conclusions.

Throughout this study, we describe the spectral and temporal evolution of the data using the convention by which F$_\nu\propto t^{-\alpha}\nu^{-\beta}$. We adopt a cosmological model with $H_0=67.3\,\textnormal{km\,s}^{-1}\textnormal{Mpc}^{-1}$, $\Omega_M=0.315$, $\Omega_\Lambda=0.685$ \protect\citep{Planck2014}.

\section{Observations}\label{sec:observations}

\subsection{High-Energy data}

GRB\,210610B was first detected by the Gamma-ray Burst Monitor (GBM) onboard the \textit{Fermi} observatory at 19:51:05.05 UT of 10 June 2021 with a T$_{90}$ = 55.04$\pm$0.72 s and a fluence of $(1104.2\pm0.5)\times10^{-5}$\,erg\ cm$^{-2}$ in the 10\,keV to 10\,MeV band \citep{MalacariaGCNFermiGBM, vonKienlin2020FermiOnlineCat}. At the redshift of the GRB (see sect. \ref{sec:spectrum}), the computed isotropic energy is $E_{\textnormal{iso, rest}} =  4.17_{-0.02}^{+0.02}\times10^{53}$\,erg in the 0.1\,keV to 10\,MeV band. This value is fully consistent with the so-called ``Amati'' relation \citep{Amati2002AA, Amati2006MNRAS} for long GRBs. The burst was detected $\sim$22\,s later by the Burst Alert Telescope (BAT, \citealt{Barthelmy2005SwiftBAT}) on board the \textit{Neil Gehrels Swift Observatory} with coordinates $\textnormal{RA}=16^\textnormal{h}\;15^\textnormal{m}\;45^\textnormal{s}$, $\textnormal{Dec.}=+14^{\circ}\;23\arcmin\;29\arcsec$ and an uncertainty of {3\arcmin}. The X-Ray Telescope (XRT, \citealt{Burrows2005SwiftXRT}) started observing the source 89.9 s after BAT and quasi-simultaneously, the Ultraviolet/Optical Telescope (UVOT, \citealt{Roming2005SwiftUVOT}) observed the field in the \textit{White}-band. The image showed a new bright source within the XRT position with a magnitude in the \textit{Swift}/UVOT native system of 13.70$\pm$0.14, uncorrected for galactic extinction \citep{Page2021BATDetection}. The later analysis, presented in \cite{SiegelGCNUVOT}, updated this value to 13.63$\pm$1.10 and the source location to $\textnormal{RA}=16^\textnormal{h}\;15^\textnormal{m}\;40.40^\textnormal{s}$, $\textnormal{Dec.}=+14^{\circ}\;23\arcmin\;56.9\arcsec$ with an uncertainty of {0\farcs42}.

In the refined analysis from the BAT data, the burst had a $T_{90}$ duration of T$_{90}$ = 69.38$\pm$2.53\,s in the 15 to 350 keV band with an $E_{\rm peak}$ = 339.3$\pm$218.6 keV\citep{KrimmGCNRefinedBAT}. The burst showed a hardness ratio HR\footnote{The hardness ratio or spectral hardness \citep{Kouveliotou1993} for a GRB is the ratio between the fluence emitted during the prompt-emission in the 50--100 keV band over the 25--50 keV band.} of 1.78$\pm$0.04. This falls in the upper part of the bulk of the HR distribution for long GRBs \citep[see e.g.][]{Lien2016}. The burst was also detected by Konus-\textit{Wind} observatory with a duration $\sim$ 100\,s \citep{FrederiksGCNKonusWind}. The HR of the burst, its T$_{90}$ and its isotropic energy release classifies GRB\,210610B as a long GRB.

\subsection{Linear polarimetry imaging}

We observed the GRB\,210610B following the \textit{Swift}/BAT alert with two different instruments in linear polarimetry mode. We first observed with the Calar Alto Faint Object Spectrograph (CAFOS) mounted on the 2.2 meter telescope at Calar Alto Observatory\footnote{Observations were obtained with programme F21-2.2-021 (PI: Ag\"u\'i Fern\'andez, J. F.).}. The observations started 0.08 days after the \textit{Swift}/BAT detection with an exposure per half-wave plate (HWP) position angle (see below) of 900\,s in \textit{R}-band. A second observation was obtained with the FOcal Reducer and low dispersion Spectrograph (FORS2) mounted on the Unit Telescope 1 of the Very Large Telescope (VLT) of the European Southern Observatory (ESO)\footnote{Observations were obtained with program 106.21T6.003 (PI: Tanvir, N.).}. The VLT/FORS2 observations started 0.24 days after the GRB alert in \textit{R$_{\text{Special}}$, b$_{\text{High}}$} and \textit{z$_{\text{Special}}$}-bands with 180\,s exposure time per image and with two cycles with the same configuration.

For both instruments observations were obtained using the HWP in four different rotation angles of 0.0$^{\circ}$, 22.5$^{\circ}$, 45.0$^{\circ}$ and 67.5$^{\circ}$ and a Wollaston prism to split the light into the ordinary (\textit{o}) and extraordinary (\textit{e}) beams. A mask was set to avoid overlapping of the light from both beams.

For calibrating the CAHA/CAFOS observations, one high- and one zero-polarised standard stars were observed following the same procedure as for the GRB. Flat-fields in the corresponding band were obtained with the full optical setup in the light path. For the VLT/FORS2 data, calibration and standard observations where performed as specified in the FORS2 User Manual \citep{FORS2manual}.

We obtained a second epoch with VLT/FORS2 in \textit{R$_{Special}$} $\sim$1 day. We increased the exposure time per  image up to 300\,s to account for the fading of the source performing two cycle for this observations.

\subsection{Photometry}

Photometric observations were performed using several instruments at different observatories and multiple bands. In this section we give details of each of the observations. The measured photometry is compiled in Table \ref{Tab:phot}.

\subsubsection{Small Binocular Telescope}
\label{sec:sbt}

The Small Binocular Telescope (SBT), located at the Ond\v{r}ejov Observatory, observed GRB\,210610B in dusk with the two 20\,cm Newtonian astrographs mounted on a common mount. The main detectors have 4096$\times$4096 pixel CCDs that provide a field-of-view of 3.5$^\circ\times$3.5$^\circ$ with a 3.14$"$ sampling. Operations are designed such that the readout time of one camera equals the exposure time of the other one to avoid any blind time during the  observation.

The SBT observations started $\sim$\ 860\,s after  burst. During the early follow-up phase, 12\,s exposures were taken without a filter. Afterwards, the observations consisted of up to 34 exposures of 120\,s each in SDSS-\textit{r'} band (see Table~\ref{Tab:phot} for further details). Observations were interrupted at 23:47\,UT, almost 4\,h after the GRB detection.

The afterglow was not detected in single exposures so we stacked the images 
to obtain an acceptable signal-to-noise ratio. 
We used \texttt{montage} to obtain a weighted image co-addition similarly to \cite{Morgan2008} which optimises for the variable background caused by both, dusk and thick cirrus. 

As for the images in \textit{r'}-band, photometric calibration was performed against the Atlas catalogue that uses Pan-STARRS catalogue for faint targets. Images in the clear band were calibrated against \textit{r'}-band and a polynomial correction involving photometric colours $g'-r'$, $r'-i'$ and $i'-z'$. Photometric measurements obtained this way are shifted compared to the original standard \textit{r'}-band by a correction that depends on the object colour. This correction can be computed from the fitted relation if the photometric colours of the afterglow are known. By fitting the complete photometric set with an empirical broken power law we determined this correction to be $k_\mathrm{c-r} = -0.0045$\,mag. Therefore, under the assumption of no colour evolution of the afterglow, we are able to convert the unfiltered values from SBT to $r'$ by simply subtracting $k_\mathrm{c-r}$.

\subsubsection{FRAM-ORM telescope}
\label{sec:fram}

The 25\,cm telescope FRAM-ORM, is operated as part of CTA-N at Roque de los Muchachos at La Palma, Spain. The telescope is equipped with a $1024\times1024$ pixel CCD detector and a Bessel filter set, which provides a field of view of 26$'\times$26$'$ with 1.5$"$/pixel scale.

FRAM-ORM observed the GRB starting at 20:49\,UT (1h after trigger) and obtained a total of 83$\times$60\,s images in \textit{R}-band using a total of 1.5\,h of observing time. Images were reduced using standard procedures and combined in groups in order to provide a good signal-to-noise ratio. Photometry was performed following a similar fashion to SBT unfiltered images, using \textit{r'}-band and \textit{r'-i'}-band catalogue values. The derived correction factor is $k_\mathrm{R-r} = -0.036$ mag. 

\subsubsection{Telescope D50}
\label{sec:d50}

The D50 Telescope is a 0.5\,m Newtonian robotic telescope located at Ond\v{r}ejov Observatory. It has a 1024x1024 EMCCD camera with a field-of-view 20$'\times$20$'$, scaled at 1.18$"$ per pixel. The telescope is equipped with an SDSS filter set.

D50 started observing 0.06 days after the burst detection, starting  with SDSS-\textit{r'}-band and followed by a set of observations in SDSS-\textit{g'-}, \textit{r'-}, \textit{i'-} and \textit{z'-}bands using different exposure times (see Tab.\ref{Tab:phot} for further details). During the first night, the telescope spent 3.2\,h on target and stopped at 01:05\,UT.  There were additional observations at night 2, 4, 5 and 7,  collecting 16.75\,h of further follow-up data.

All images were processed in a standard manner performing dark subtraction, flat-field correction and fringe removal for $i'$ and $z'$ and the images were co-added when necessary. Photometry was performed using the SDSS catalogue in the corresponding band. The GRB afterglow emission is detected even at the latest epoch. 

\subsubsection{CAHA/CAFOS}
\label{sec:cafos_photometry}

In addition to the polarimetric observation, we performed further imaging of the afterglow on the second night with CAFOS on the 2.2\,m telescope of the Calar Alto Observatory. The images were corrected using bias and flat fields using standard procedures in IRAF. The observations were performed with the $R_C$ filter and calibrated with respect to PanSTARRS field stars, for which a filter correction was used to derive AB magnitudes in the $R_C$ band. These magnitudes were then transformed to the $r'$ band using the colour information that we have from the afterglow. The final values are shown in Table~\ref{Tab:phot}.

\subsubsection{GTC/HiPERCAM}
\label{sec:hipercam}

The field of GRB\,210610B was observed in 4 epochs using the HiPERCAM multi-band imager \citep{Dhillon2021} mounted on the 10.4 m Gran Telescopio de Canarias (GTC) at Roque de los Muchachos Observatory (La Palma, Spain) using programme GTCMULTIPLE2C-21A (P.I.: de Ugarte Postigo). HiPERCAM simultaneously obtains observations in the five SDSS filters (\textit{u'}, \textit{g'}, \textit{r'}, \textit{i'}, \textit{z'}) using efficient dichroic beam-splitters and multiple cameras. The last of our observations was obtained almost 2 months after the burst when the emission was dominated by the host galaxy (see Fig. \ref{Fig:FC}).

The data reduction was performed using an automatic shell script that finds and organises the files, calls commands from the HiPERCAM pipeline to perform bias and flat corrections and converts the HiPERCAM one-dimension fits files to classical two dimension fits images. Further IRAF procedures allow to obtain an even background from the different quadrants of each detector. Finally, the images are registered and combined using SWARP \citep{Bertin2010}. Photometry was performed with aperture photometry using reference field stars from the PanSTARRS catalogue. For the last epoch we used the same aperture to compare the results to the rest of the data. Additionally, we performed photometry of the complete host galaxy adapting the aperture to its light, and in a similar way we obtained photometry of a nearby object north-west of the host, which we identify as a companion galaxy at the same redshift (see Sect. \ref{sec:host}).

\subsubsection{Perek 2\,m telescope}
\label{sec:perek}

The Perek 2.0\,m telescope at the Ond\v{r}ejov Observatory observed GRB\,210610B afterglow 6 days after the burst in SDSS-\textit{g'} band. The photometric camera of this telescope has a 1092$\times$736 pixel CCD with a field of view of $5' \times 7'$ scaled at 0.4$''$/pixel. After standard imaging data reduction and after imaging co-add, photometry was performed as for D50 telescope. The GRB afterglow is well detected.

\subsubsection{GTC/EMIR}

Late time near-infrared (NIR) observations were performed on 19 February 2022 in search for the host galaxy of GRB\,210610B with the EMIR instrument \citep{Garzon2022} mounted on the 10.4m GTC telescope with programme GTCMULTIPLE2H-21B (PI: de Ugarte Postigo). The observation consisted of a total exposure of $349\times3$ s in $H$-band. The data reduction was performed using a self-made pipeline that corrects flat fields, does background subtraction, bad pixel masking, alignment and combination of the frames. The host galaxy was not detected in the final frame down to a 3-$\sigma$ limiting magnitude of 22.9 mag.

   \begin{figure}
   \centering
   \includegraphics[width=\hsize]{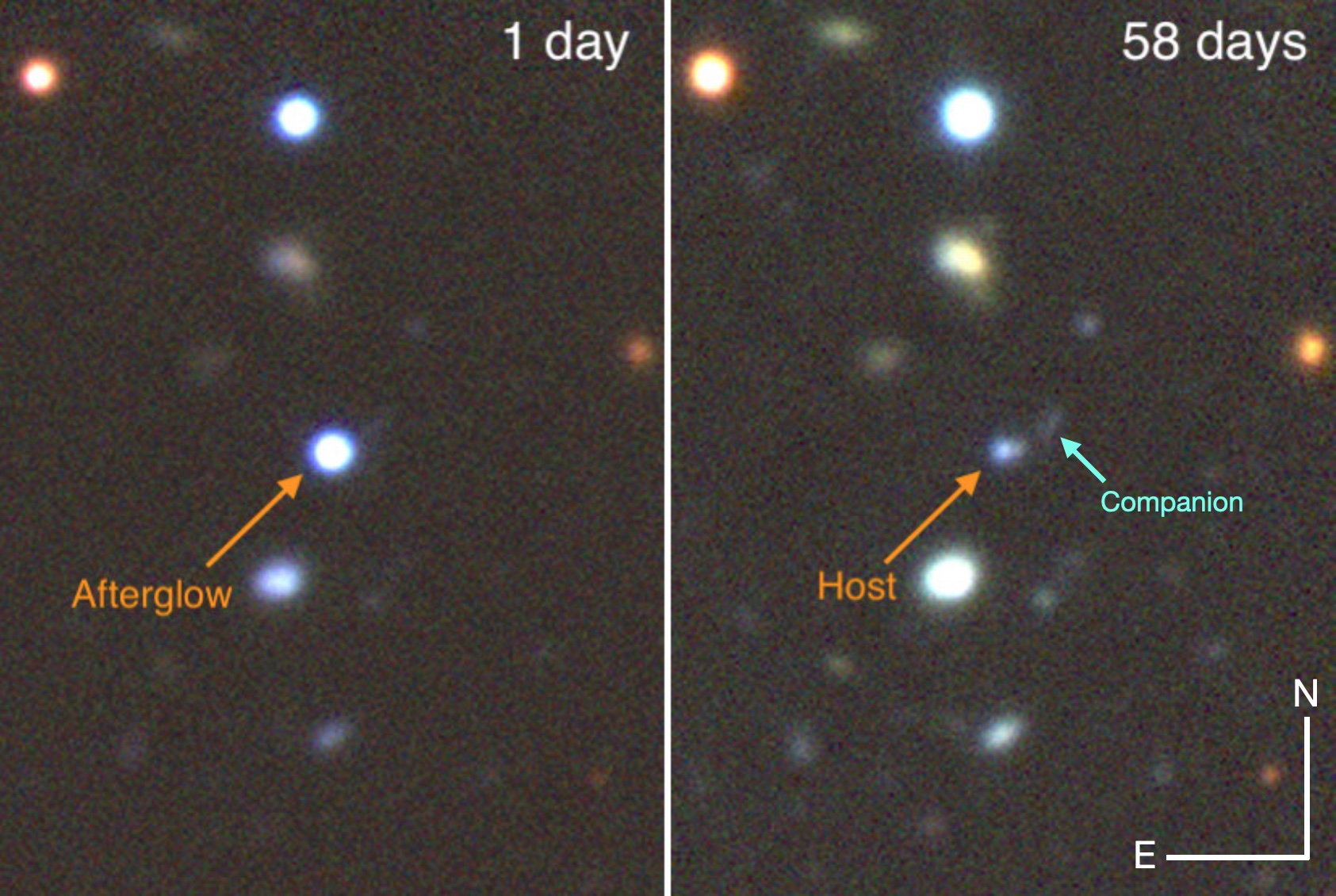}
      \caption{Colour images of the field using the \textit{g'}, \textit{r'} and \textit{i'} bands of HiPERCAM at 1 day and 58 days after the burst, respectively. In the first image the afterglow is strongly detected, in the second one it has faded and the host galaxy dominates the emission. 2.7\,arcsec North-West of the host galaxy is a companion galaxy at the same photometric redshift of the host (see Sect.\ref{sec:host}).
              }
         \label{Fig:FC}
   \end{figure}

\subsection{Spectroscopy}

We observed GRB\,210610B using the longslit mode of the Optical System for Imaging and low Resolution Integrated Spectroscopy (OSIRIS) \citep{OSIRIS2000SPIE} mounted on the 10.4~m GTC. The observation consisted of 3$\times$900~s exposures with grism R1000B and a slit width of 1$^{\prime\prime}$ oriented at the parallactic angle\footnote{The observations were obtained under the program GTCMULTIPLE2C-21A (PI: de Ugarte Postigo).}. The mean epoch of the observation was 11 June 2021 at 01:59:14 UT (6.12972 hrs after the \textit{Swift} trigger) at a mean airmass of 1.13.

Flux calibration was performed relative to an observation of the Ross\ 640 \citep{Oke1974} spectrophotometric standard observed at the beginning of the same night and using the same grism. The afterglow spectrum shows a very strong continuum, with a median signal-to-noise ratio of $\sim100$ per dispersion element and weak absorption features.

\section{Results}\label{sec:results}

\subsection{Linear polarimetric analysis}
\label{sec:polarimetry_analysis}

For all linear polarimetry  data we performed regular imaging reduction processes. Images were bias subtracted and flat-field corrected using \texttt{PyRAF} tasks \citep{pyraf}. The flat-field correction for VLT/FORS2 was performed as specified in \cite{Santiago2020}. For CAHA/CAFOS, we combined the flat-fields using \texttt{PyRAF} and used a customized \texttt{Python} script to separate the \textit{o} and the \textit{e} beams into two separate images. We normalised each beam to the corresponding median value and created the  \textit{o} and  \textit{e} beam  combined flat. We followed the same procedure for each image. We did this correction since flat-fields were obtained with the full optics on the light pathway. This led to a rather different count rate in each beam what would have led to an non-accurate normalisation. Finally, we split all the reduced images into an \textit{o} and \textit{e} image and applied the \texttt{L.A. Cosmic} algorithm \citep{VanDokkum2001} for cosmic ray removal.

We used \texttt{PyRAF} to measure the on-frame full width at half-maximum (FWHM) in each image  including the high- and the zero-polarised standard stars. The FWHM was measured independently for the \textit{o} and \textit{e} beam since the shape of the point spread function (PSF) can vary, especially for sources with a high degree of polarisation.

To obtain reference field stars we used a source detection algorithm based on \texttt{DAOStarFinder} in \texttt{Photutils} and applied it  to the background subtracted \textit{o} and \textit{e} images separately. The selected sources were those with threshold above the median plus three times the standard deviation of the background. The statistics were calculated per beam and per angle position of the HWP using a sigma clipping of the masked image. From these sources, we discarded those that showed to be clearly extended and those too close or partially within the instrumental mask edges. We also checked if the source was saturated in any of the beams. We ended up with 5 sources, including GRB\,210610B, in the FORS2 images and 5 in the CAFOS ones.  Sources from FORS2 and CAFOS images are different due to saturation or sources falling in the mask edges.

We performed aperture photometry using circular apertures with a radius equal to the FWHM and applied infinite-aperture corrections using a self-made \texttt{Python} script to the FORS2 data. As for the CAHA/CAFOS data, in order to avoid contamination by a nearby spurious source, we measured the flux using a fixed aperture of 3 times the FWHM per image and per beam and subtracted the sky of an annulus around the source with an inner and outer radius of 4 and 5 times the FWHM, respectively. In the FORS2 images, this spurious source was outside the  measured region. 
The errors for the aperture photometry were obtained considering each beam as an unique image, i.e., we separated the ordinary stripes from the extraordinary ones.

With the measured flux values per source and per beam, \textit{f$_{o}$} and \textit{f$_{e}$}, we obtained the Stokes parameters \textit{Q} and \textit{U} describing the linear polarisation (see e.g. \citealt{Patat2006, Bagnulo2009}) for each of our images. We used the normalised Stokes parameter for linear polarisation.

\begin{eqnarray}
   \dfrac{Q}{I} = q = \dfrac{2}{N} \sum_{i=0}^{N-1} {F_i} {\cos{\dfrac{i\pi}{2}}}\\
   \dfrac{U}{I} = u = \dfrac{2}{N} \sum_{i=0}^{N-1} {F_i} {\sin{\dfrac{i\pi}{2}}}
\end{eqnarray}

where \textit{N} is the number of positions of the half-wave plate, \textit{I} is the intensity and \textit{F$_{i}$} is the normalised flux difference per HWP position angle,

\begin{eqnarray}
   F_{i} = \dfrac{f_{o, i} - f_{e, i}}{f_{o, i} + f_{e, i}}
\end{eqnarray}

Following the equations above, we can obtain the polarisation degree \textit{P$_{lin}$},

\begin{eqnarray}
   P_{lin} = \sqrt{q^2 + u^2}\\   
\end{eqnarray}

For the position angle, we followed the formalism as used in \cite{Bagnulo2009}.

We then corrected the effects of optics and detector on the polarisation images. To do so, zero-polarised standard stars were observed in the corresponding bands with VLT/FORS2, following the procedures detailed in \cite{FORS2manual}. For CAHA/CAFOS, observations of standard stars were carried out on the same night. In the case of VLT/FORS2, \textit{b-} and \textit{z-}band standards were completely saturated and no subsequent standards were found in the ESO archive around the time of the observation. For the \textit{R-}band, the standards were also saturated. We then utilised WD1620-391 for the zero polarised standard star, observed 18 days after the GRB, for VLT/FORS2, and BD+33\,2642 \citep{Turnshek1990} for CAHA/CAFOS  observed the same night as GRB\,210610B.

\begin{figure}[!ht]
\centering
\includegraphics[width=\hsize]{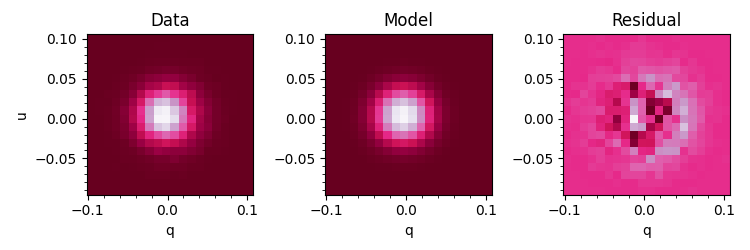}
   \caption{\textit{Left}: Normal distribution of randomly generated data around \textit{q, u} values with the amplitude to generate the data is Stokes parameter errors. \textit{Central panel:} 2 dimensional gaussian model fitted to the data points. \textit{Right:} Residuals.
           }
      \label{Fig:2d_gaussian}
\end{figure}

We finally removed the effect on the polarisation induced by the dust in the Milky Way (MW). For this, we measured the \textit{q} and \textit{u} parameters for the field stars in our images. However, FORS2 is known to have a radial profile with polarisation varying across the field from the optical axis towards the edges of the detectors \citep[see e.g.][]{Santiago2020}. To account for this effect, we applied the corresponding correction by using the \textit{q, u} background correction and the instrumental polarisation maps presented in \cite{Santiago2020}. We then measured the Galactic interstellar polarisation (Galactic ISP or GISP) using three methods, following \cite{Wiersema2012}: First, we measured the mean values for the \textit{q, u} parameters from the field stars. Then we performed a 1-Dimensional Gaussian fit to the \textit{q, u} values of the field stars by generating a normal distribution of values for the \textit{q, u} within their corresponding errors and then fitted a Gaussian to each Stokes parameter. Finally,  we performed a 2-Dimensional Gaussian fit to the \textit{q, u}, adapting the procedure from the 1-Dimenssional Gaussian fit (see Fig. \ref{Fig:2d_gaussian} for an example of the 2D Gaussian fit). All the \textit{q, u} fitted parameters are consistent within errors so we  vectorially subtracted the mean value from the \textit{q, u} values for the GRB. The mean sky values are listed in Tab. \ref{table:plin_theta_values}, which are all consistent or close to zero. Both \textit{q} and \textit{u} show some variation from one epoch to the next, which we assume are due  different sky background light between epochs\footnote{\href{https://www.eso.org/asm/ui/publicLog?name=Paranal&startDate=2021-06-10T21:00:00.000Z&hoursInterval=15}{ESO weather log}.}.

For the CAFOS images, the same analysis was performed although we could not correct the background polarisation and instrumental polarisation across the images given the lack of a characterisation as we have for FORS2. However, lunar illumination was close to 0\% and we do not expect a high instrumental polarisation \citep{Patat2011CAFOS} for CAFOS. Indeed, the 2-D Gaussian fit to the field stars (see Fig. \ref{Fig:2d_gaussian}) shows it to be consistent with no polarisation  from the ISM and, implicitly, from the CAFOS instrument.

After the polarisation induced by the MW dust was removed we calculated the final polarisation from both instruments in all the filters. Next, we needed to consider the contribution to the polarisation from the dust in the host galaxy itself. Since the information is more limited than for the dust in the MW we assumed a Serkowski law \citep{Serkowski1975} for the host galaxy. From the Spectral Energy Distribution (SED) fit of the GRB light-curve we obtain the color excess on the line-of-sight (see Sect. \ref{sec:light-curve}). Assuming this value as the extinction of the afterglow at the host galaxy, using P$_{\textnormal{ISP}}(\%) \leq 9.0 \times E(B-V)$, we find the contribution from the host galaxy could be as high as $P_{\textnormal{Host ISP}} \leq 0.09 \%$. We note that in extragalactic sight-lines, this may not be applicable since dust properties may differ from those of the MW \citep{Nagao2022}. This value is well below the PD measured for the afterglow, so we consider the host contribution to be negligible.
Finally, we corrected for the polarisation definition bias with the Modified ASymptotic Estimator (MAS) \cite{Plaszczynski2014}. The corrected values for the polarisation degree are shown in Tab.\ref{table:plin_theta_values} before and after the bias correction. 

We also determine the polarisation position angle (PA) for the GRB afterglow. The FORS2 measured raw PA was corrected for chromaticity using the tabulated values per bandpass presented in \cite{FORS2manual}. We then corrected it using the standard star Hiltner 652, observed on the 19$^{th}$ of July, 2021, although the PA correction is very small with $\Delta \theta = 0.5 \pm 0.9$. The high polarisation standard in \textit{R$_{Special}$}-band closest in time to the GRB was also saturated in at least one beam. We corrected the derived PA value of the standard to the one presented in \cite{Cikota2017}. As the measured \textit{b-}band value for the PA in Hiltner 652 is very close to the \textit{B-}band value in \cite{Cikota2017} we  use \textit{B}-band to correct the \textit{b-}band value. For the \textit{z-}band,  no observations has been found so far for this standard, hence we used the PA for \textit{I-}band in \citep{Cikota2017}. As for CAFOS, the high polarisation standard observed, Hiltner 960, was corrected to the ``theoretical'' value following \cite{Schmidt1992}. 

We detect polarisation at $\gtrsim$3$\sigma$ significance for the first and the last epoch while the second epoch is consistent with zero polarisation. We detect 1-$\sigma$ polarisation in \textit{R-}band on the  second epoch. However we consider this measurement, together with the \textit{R}-band observation at $\sim$0.26\ days, \textit{b-} and \textit{z-}band as limit. Since the MAS correction is not completely applicable when PD/$\sigma_P$ $<$ 3.0, we do not apply this correction to the mentioned limits.  In this PD regime, the PA would behave erratically and, therefore, we cannot treat it as a limit.

\begin{table*}
\begin{center}
\caption{Measured values for the linear polarisation and PA of GRB\,210610B. When the standard corrected PA is negative, it has been corrected to positive values by subtracting it from 360º. We measured the signal-to-noise ratio (S/N) in each beam and at each HWP angle image. The value we present here is the lowest we measure from in one beam in one image at a certain HWP angle.}
\label{table:plin_theta_values}
\begin{tabular}{ccccccccc}
\hline\hline  
Epoch & Bandpass & Instrument/Telescope & q$_{sky}$ & u$_{sky}$ & P$_{Lin}$ & P$_{Lin, Debiased}$ & $\theta$ & S/N \\ 
 t$-$t$_0$\,(day) &  &  & (\%) & (\%) & (º) & \\ 
\hline 
0.1205 & \textit{R$_C$} & CAHA/CAFOS & 0.2$\pm$2.4 & -0.01$\pm$0.31 & 4.50 $\pm$ 1.45 & 4.27 $\pm$ 1.45 & 183 $\pm$ 9 & > 500 \\
0.2418 & \textit{R$_{Special}$} & VLT/FORS2  & -0.30$\pm$0.32 & -0.29$\pm$0.20 & 0.28 $\pm$ 0.20 & -- & 267 $\pm$ 19 & > 710 \\
0.2593 & \textit{R$_{Special}$} & VLT/FORS2  & -0.79$\pm$0.28 & 0.81$\pm$0.22 & 0.60 $\pm$ 0.24 & -- & 17 $\pm$ 11 & > 690 \\
0.2698 & \textit{b$_{High}$}    & VLT/FORS2  & -- & -- & 0.18 $\pm$ 0.16 & -- & 187 $\pm$ 24 & > 395 \\
0.2803 & \textit{z$_{Special}$} & VLT/FORS2  & -- & -- & 0.23 $\pm$ 0.28 & -- & 199 $\pm$ 35 & > 180 \\
1.2674 & \textit{R$_{Special}$} & VLT/FORS2  & -0.13$\pm$0.22 & 0.06$\pm$0.07 & 2.28 $\pm$ 0.22 & 2.27 $\pm$ 0.22 & 237 $\pm$ 3 & > 330 \\
1.2766 & \textit{R$_{Special}$} & VLT/FORS2  & -0.36$\pm$0.39 & 0.06$\pm$0.07 & 1.72 $\pm$ 0.27 & 1.69 $\pm$ 0.27 & 238 $\pm$ 5 & > 300 \\
\hline 
\end{tabular}
\end{center}
\end{table*}

\begin{figure}[!ht]
\centering
\includegraphics[width=\hsize]{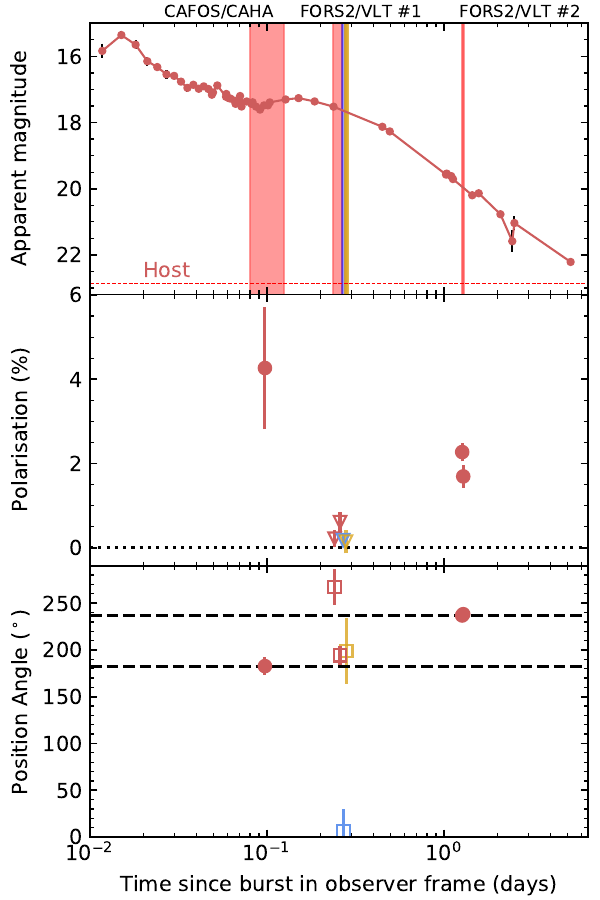}
   \caption{\textit{Top panel:} GRB\,210610B light-curve for the \textit{r-}band (see Tab. \ref{Tab:phot}). We convert the SBT \textit{Clear-}band and \textit{R-}band to \textit{r-}band as indicated in \ref{sec:sbt} and \ref{sec:fram}. The vertical stripes, from left to right, denote  the first polarimetry epoch (CAHA/CAFOS) and the second and third epochs (both with VLT/FORS2). The host observations in \textit{r-}band are marked with a red dashed line. \textit{Middle panel:} GRB\,210610B linear polarisation evolution. Red dots shows the measured polarisation for \textit{R}-band while  blue and orange triangles denote \textit{b} and \textit{z}-band lower limits. \textit{Bottom panel:} Linear polarisation measured PA for GRB\,210610B , detections are marked with filled circles while the corresponding PA for PD limits are marked with empty squares as this values are neither lower nor upper limits. Black dashed lines denote the PA for the first (bottom) and the third (top) epoch.}
      \label{Fig:polarisation_plot}
\end{figure}

\begin{figure*}[!ht]
\centering
\includegraphics[width=0.8\textwidth]{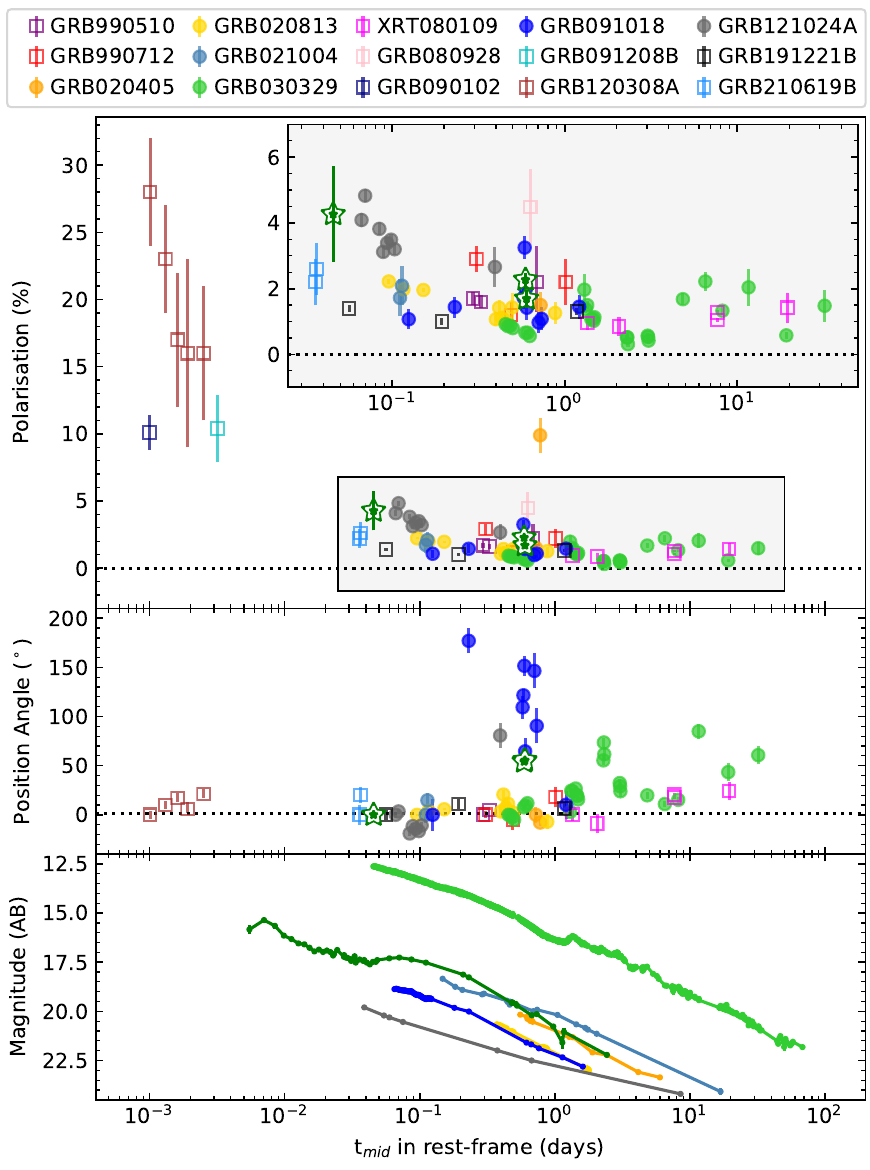}
   \caption{\textit{Top: }Measured GRB linear polarisation degree on the optical afterglow emission. The white and green stars marks the polarisation degree measured for GRB\,210610B in all bands. As for all the data-points, we do not make a distinction on the photometric band in which polarisation was measured. We selected only the measured values that shows a P/$\sigma_P$ $>$ 3 for all bursts, including GRB\,210610B. We show in filled circles those burst for which a light-curve is shown in bottom panel, empty squares are those that are not represented in this last panel. Data and references can be found in Tab. \ref{table:plin_theta_values_comparison}. \textit{Middle pannel: }PA measured for each corresponding burst and epoch in the same fashion as on the top panel. Note that as for GRB\,021004 there is no measured PA. To better distinguish PA changes, we subtracted the first PA value to all values and calculated the absolute value for those with a mean value below zero. We do not find significant PA changes except for GRB\,121024A, GRB\,091018, GRB\,030329 and GRB\,210610B. We note that the measures are carried out with different instruments and data reduction and analysis can be different from the one we follow in this work. \textit{Bottom panel: } We show the light-curve for some exemplary burst with data available in literature (GRB\,020405 \citep{Masetti2003}, GRB\,020813 \citep{Gorosabel2004}, GRB\,021004 \citep{deUgartePostigo2005}, GRB\,030329 \citep{Lipkin2004}, GRB\,091018 \citep{Wiersema2012}, GRB\,121024A \citep{Wiersema2014} and GRB\,210610B.
           }

      \label{Fig:comparison_polarisation}
\end{figure*}

\subsection{Light-Curve analysis}
\label{sec:light-curve}

\begin{figure}[!ht]
\centering
\includegraphics[width=\hsize]{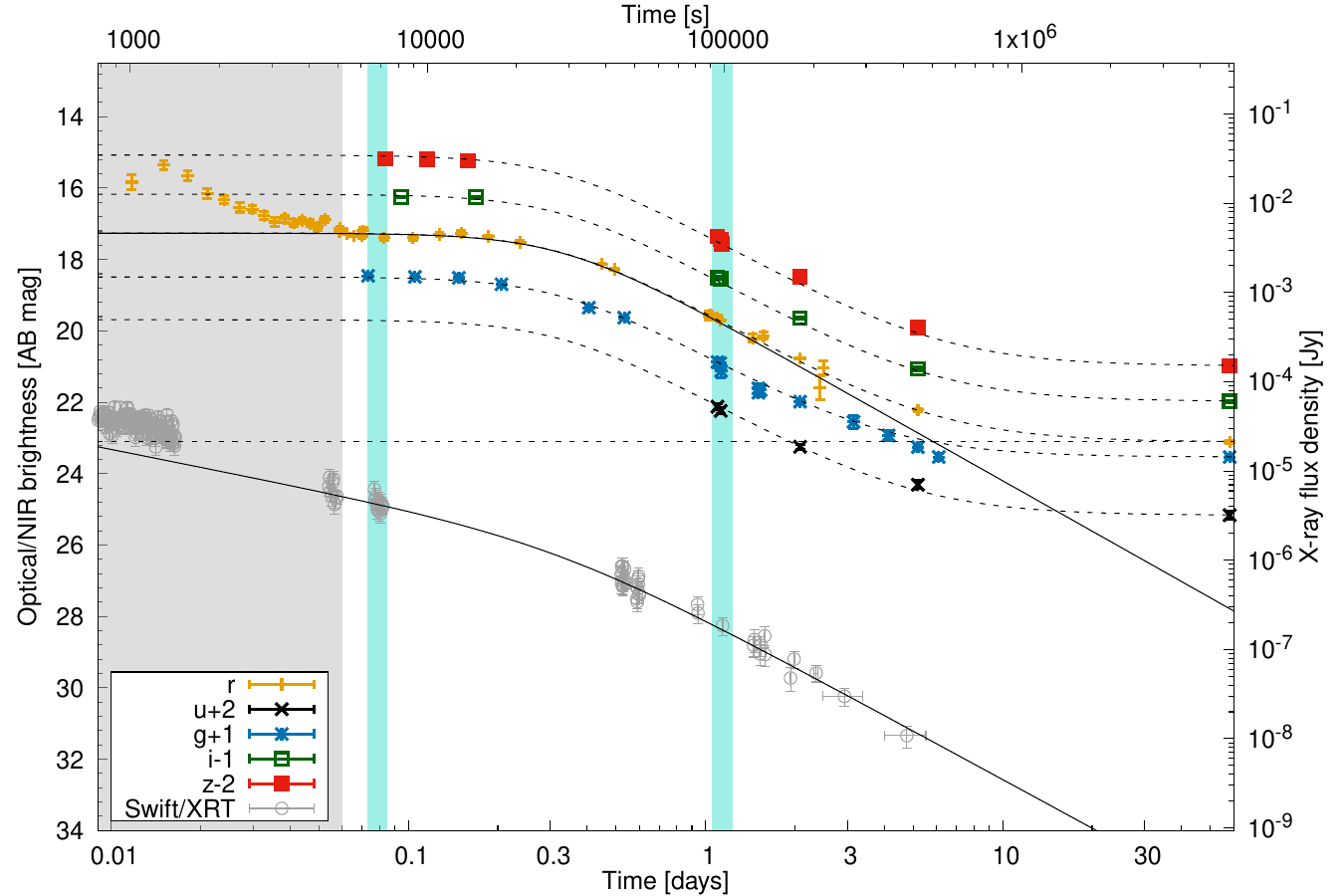}
   \caption{GRB optical to X-ray light curve fit to available photometry (see Tab. \ref{Tab:phot}) and \textit{Swift} X-ray data. The gray-shaded region is not taken into account for the fit as it may be contaminated by the prompt emission. Clear  cyan regions mark the times we choose to derive the SED of the light-curve. Data are shown with an offset in flux for better visibility.
           }
      \label{Fig:lc_SED_fit}
\end{figure}

We first modelled the optical and X-ray light curves simultaneously with a smoothly broken power-law \citep{Beuermann1999a}:
$F = (F_1^{\kappa}+ F_2^{\kappa})^{-1/\kappa}$,
where $F_\textrm{x}=f_\textrm{break}(t/t_\textrm{break})^{-\alpha_x}$,
$f_\textrm{break}$ being the flux density at break time $t_\textrm{break}$, $\kappa$ the break smoothness parameter, and the subscripts $1,2$ indicate pre- and post-break, respectively.
We did not consider data before 0.06 days ($\sim$5 ks) since they are still dominated by the initial rapid decay.
Even before performing any modelling, one can clearly see that the optical light-curve is initially flat at least until $\sim$0.25\ days ($\sim$20\ ks, see Fig. \ref{Fig:lc_SED_fit}), while at the same time the X-ray behaviour is difficult to discern, though the two groups of observations at 4 ks and 7 ks show possible fading, in contrast to the optical.
Not knowing the precise evolution of the early X-ray data, we allow the initial X-ray decay to be different from the optical. Note that a different decay implies a colour evolution between X-rays and optical.
We find a shallow break with $t_\textrm{break,~opt} = 0.326\pm 0.011$ d ($27.7\pm1.2$~ks), a flat optical decay with $\alpha_{1,\rm opt}=0.00\pm0.01$, and an  optical to X-ray  decay index $\alpha_{2,\rm}= 1.85 \pm 0.04$, with break smoothness $\kappa=1.4\pm0.1$, with  $\chi^2/\mathrm{d.o.f.}$=2.6.

After the break we analysed the full optical to X-ray SED when we have the best coverage in both frequency regimes. To build this SED, we first created the XRT spectra using the time-slice tool in the XRT repository \citep{Evans2007a,Evans2009a} in the range 30ks-50ks at mid-time of 1.2 days. We then shifted the optical GTC/HIPERCAM ugriz data at 1.097 days closer to these times using the decay indices found above.

We modelled the afterglow SED from optical to X-ray frequencies using \texttt{Xspec v12.13.0} \citep{Arnaud1996}. The redshift was fixed to $z=1.1341$ and we fixed the Galactic hydrogen column density to
$N_H=3.94\times10^{20}\, \textnormal{cm}^{-2}$ \citep{Willingale2013a}.
The data is best modelled by a single power-law ($\chi^2/\mathrm{d.o.f.}=264.5/298$) with $\beta=0.869_{-0.007}^{+0.003}$, intrinsic absorption $N_H=18.3_{-5.9}^{+6.5}\times10^{20}\, \textnormal{cm}^{-2}$ (using the Tuebingen-Boulder ISM absorption model; \citealt{Wilms2000}) and E(B-V)\,$<$\,0.01 mag, using the Small Magellanic Cloud (SMC) extinction law \citep{Pei1992a}.

We also obtained an optical to X-ray SED at 0.079 days (6871 s) taking the X-ray data between 6000 and 8000 s and using the optical light-curve above to shift the corresponding griz photometry. After fixing the redshift, extinction and absorption as for the late epoch and following the same fitting procedure, we find that the data are best modelled by a broken power-law ($\chi^2/\mathrm{d.o.f.}=143.3/197$) with $\beta_{opt}=0.43_{-0.046}^{+0.046}$, and $\beta_{X}=\beta_{opt}+0.5$, which indicates the presence of a spectral break, like the cooling break for synchrotron emission. We note that the spectral index of the high-frequency branch is consistent with the value found at late time, suggesting that the break shifted with time to lower frequencies, which is expected for an ISM environment within the fireball model and before the jet-break happens \citep{Sari1998a}. However, the shallow decay of the early light curve cannot be explained within the standard fireball model, unless we consider a more complex scenario such as an energy injection \citep[e.g.,][]{Zhang2006a}.

\subsection{Optical afterglow spectrum}
\label{sec:spectrum}

\begin{figure}
\centering
\includegraphics[width=\hsize]{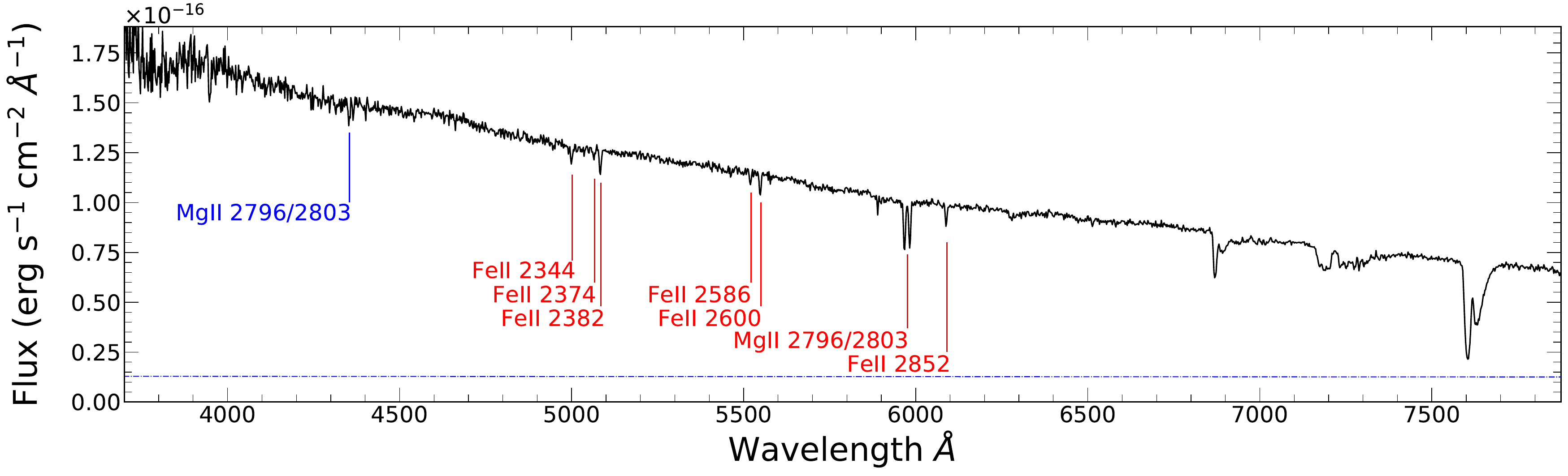}
\includegraphics[width=\hsize]{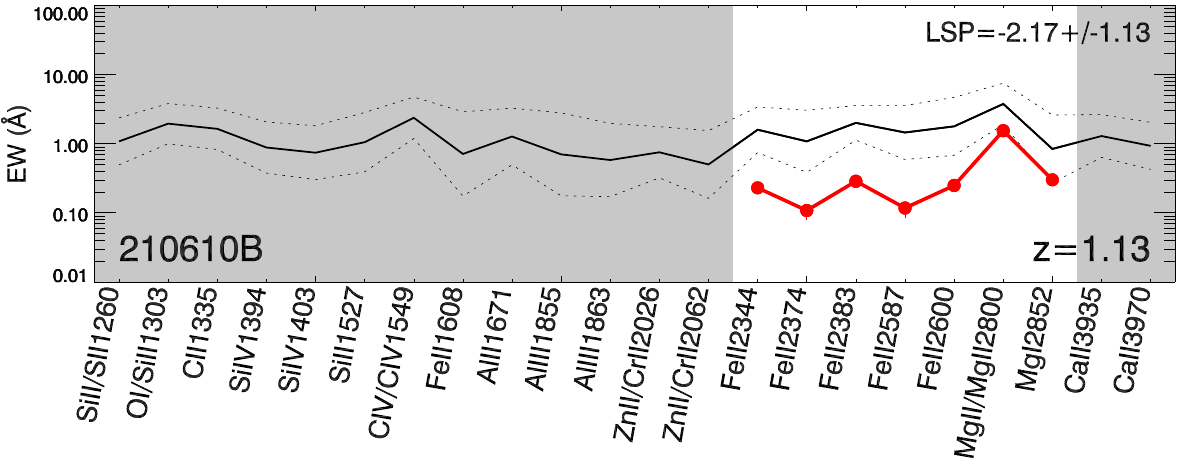}
  \caption{\textit{Top: }Red are the spectral features detected at the redshift of the GRB, blue are those from an intervening system. The dashed blue line is the error spectrum. \textit{Bottom: }Line strength diagram for the spectral features in the host galaxy. The diagram compares the features measured in our spectrum (red) with the average ones of a larger sample (black) (see Sect. \ref{sec:spectrum}).
          }
     \label{Fig:LSD}
\end{figure}

The afterglow spectrum observed by OSIRIS $\sim$0.25 days after the burst shows several transitions of \ion{Fe}{ii}, \ion{Mg}{ii} and \ion{Mg}{i} (see Tab. \ref{Tab:EWs}) all of them at a common redshift of $z=1.1341\pm0.0004$ (see Fig. \ref{Fig:LSD} and Table \ref{Tab:EWs}). This value is a lower limit for the  burst redshift due to the non-detection of fine-structure lines excited by the GRB itself. The detection of the afterglow in the bluest band from \textit{Swift}/UVOT sets an upper limit of $z=1.7$, using the so-called Lyman ``drop-out'' technique, following \citep{Jakobsson2012}. Considering this limit and the lack of absorption lines common to GRBs \citep{Christensen2011} at higher redshift, we hence adopt $z=1.1341$ as the redshift of GRB\,210610B. We also detect two absorption features that correspond to the \ion{Mg}{ii} doublet in an intervening system at $z=0.5572\pm0.0002$.

The equivalent widths (EWs) of the detected absorption lines were measured using the \href{http://grbspec.iaa.es/}{GRBspec} database tools \citep{2014SPIE.9152E..0BD, 2020SPIE11452E..18B} (see Tab. \ref{Tab:EWs}). We follow \citet{deUgartePostigo2012} to compare these values with the common trend for long GRB sight-line environments. The Line Strength Parameter (LSP) measured for GRB\,210610B is extremely low, LSP = -2.17$\pm$1.13, implying that this line of sight has weaker features than 99.85 \% of the GRBs in the aforementioned sample. The line strength diagram in Fig. \ref{Fig:LSD} shows that the Fe\,II lines are particularly weak, and are only detected thanks to the very high signal to noise ratio of the spectrum. The magnesium features are stronger, but still among the weakest in the sample. The low EW values imply a low column density and hence a low amount of gas and possibly dust in the line-of-sight consistent with a negligible dust-induced polarisation.

\begin{table}
\begin{center}
\caption{Equivalent widths in observer frame measured in the afterglow spectrum.}
\label{Tab:EWs}
\begin{tabular}{cccccc}
\hline\hline  
Observed $\lambda$ & Feature & $z^\prime$ & EW \\ 
 ({\AA}) &  &  & ({\AA}) \\ 
\hline 
5001.69 & \ion{Fe}{ii} 2344.21 & 1.1336 & 0.49 $\pm$ 0.07 \\
5066.45 & \ion{Fe}{ii} 2374.46 & 1.1337  & 0.23 $\pm$ 0.06 \\
5084.79 & \ion{Fe}{ii} 2382.77 & 1.1340 & 0.61 $\pm$ 0.07 \\
5520.91 & \ion{Fe}{ii} 2586.65 & 1.1344 & 0.25 $\pm$ 0.07 \\
5549.64 & \ion{Fe}{ii} 2600.17 & 1.1343 & 0.53 $\pm$ 0.07 \\
5976.16 & \ion{Mg}{ii} 2796.35 & 1.1341 & 3.29 $\pm$ 0.10 \\
 & \ion{Mg}{ii} 2803.53 & 1.1345 &  & & \\ 
6090.41 & \ion{Mg}{i} 2852.96 & 1.1348 & 0.64 $\pm$ 0.07 \\
\hline
4353.92 & \ion{Mg}{ii} 2796.35 & 0.5570 & 0.39 $\pm$ 0.10 \\ 
4366.46 & \ion{Mg}{ii} 2803.53 & 0.5575 & 0.23 $\pm$ 0.08 \\ 
\hline 
\end{tabular}
\end{center}
\end{table}

\subsection{Host galaxy}
\label{sec:host}

We observed GRB\,210610B location $\sim$ 58 days after the  burst using GTC/HiPERCAM and $\sim$ 253 days after with GTC/EMIR. In the HiPERCAM images we detect an underlying object at the GRB position and we consider this object to be the host. We also find a putative companion galaxy towards North-West of the host candidate (see Fig. \ref{Fig:FC}) at a distance of $\sim$2\farcs{7}, which would correspond to a distance of 22\,kpc at the redshift of the system. The photometry of the host galaxy is shown in Tab. \ref{Tab:phot} as well as the values for the companion.

We perform a SED analysis for both galaxies assuming both of them are at the GRB redshift. For this analysis, we use the SED fitting code \texttt{CIGALE}\footnote{\url{https://cigale.lam.fr/}} \citep{CIGALE2005,CIGALE2009,CIGALE2019} on its latest version. To fit the Star Formation History (SFH), we choose a delayed star-formation history with an age for the main stellar population ranging from 0.1 to 13 Gyr and a late burst with age varying from 20\,--\,500\,Myr. We allow the code to vary the corresponding mass fraction from the total galaxy mass for the late burst from 0, which implies a single decaying exponential for the SFH modelling, up to 0.6, i.e., the 60\%\ of the total galaxy mass.
We use the Initial Mass Function (IMF) as described in \cite{Chabrier2003} with a \cite{BC03} stellar population model, assuming a stellar sub-solar metallicity (\textit{Z$_\ast$}) for the galaxy (see Sect. \ref{sec:spectrum} and \ref{sec:discussion}) that can vary from 0.004 to 0.008, according to the scheme utilised in \cite{BC03}. The nebular emission is modelled considering the same metallicity values for the gas.

For the attenuation law, we consider the modified \cite{Calzetti2000} law implemented in \texttt{CIGALE}, assuming a Small Magellanic Cloud (SMC) extinction law for the attenuation of the emission lines. We do not observe a colour evolution for the Galactic-dust-corrected \citep{Schlafly11} magnitudes and we do not detect the host galaxy in the \textit{H}-band down to 22.9 magnitudes. This could be pinpointing a low level of dust emission due to low dust heating from UV massive stars photons, indicating a low number of massive stars, or an intrinsic lack of dust in this system. Since the amount of detected \ion{Mg}{}, typically formed in the explosion of massive stars, is larger compared to the amount of \ion{Fe}{} in the traced system, although still low for common lGRBs sight-lines (see Fig.\ref{Fig:LSD}), this favours the scenario of low intrinsic extinction rather than the absence of massive stars (see Sect. \ref{sec:discussion} for an extended explanation). Therefore, we choose the colour excess of the nebular lines to be lower than 0.1.
The slope of the attenuation curve \citep{CIGALE2019} is allowed to vary between -0.6 and 0.6. For the dust emission, we select the \cite{dale14} models and allow the exponent that controls the radiation field distribution of the re-emitted energy by dust heating to vary between 1.0 and 3.0.

\begin{figure}
\centering
\includegraphics[width=\hsize]{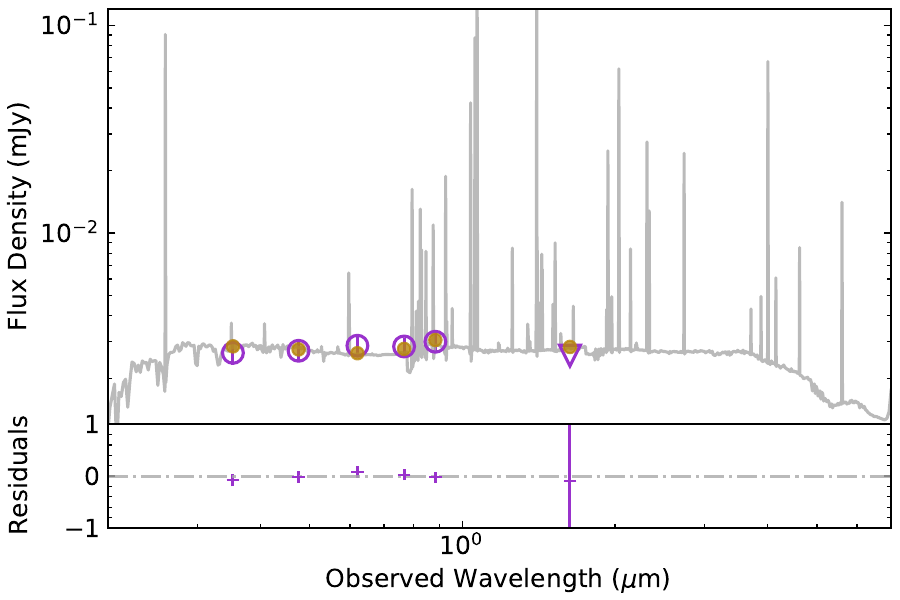}
   \caption{\textit{Top: }GRB Host Spectral Energy distribution (SED) and (\textit{bottom: }) residuals modelled using CIGALE fitting code \citep{CIGALE2005, CIGALE2009, CIGALE2019} for GRB\,210610B putative host galaxy. The vertical blue arrow indicate the \textit{H-}band upper limit.
           }
      \label{Fig:SED_host}
\end{figure}

We applied the same SED fit to both galaxies. However, the companion galaxy shows color excess in the HiPERCAM observations and therefore, we allow the colour excess for the nebular lines to range up to 0.4. 
The results for the computed physical properties from the SED fit can be found in Tab. \ref{table:host_SED_results} and the best fit is shown in Fig. \ref{Fig:SED_host}. The SED of the companion galaxy allows to determine a photometric redshift, which is consistent with the one for GRB\,210610B. The companion is at a distance of 2.67\,arcsec, corresponding to a physical distance of 22.6\,kpc at a redshift of $z=1.1341$, hence we assume these two galaxies to be part of a group.

\begin{table}
\caption{Fitted physical properties of the putative host galaxy of GRB\,210610B and its putative companion.}
\label{table:host_SED_results}
\centering
\begin{tabular}{c c c}
\hline\hline
& Host galaxy & Host companion  \\
\hline
\noalign{\smallskip}
$\log_{10}(M_{\ast}) (M_{\odot})$ & 9.10$_{-0.20}^{+0.40}$ & 9.60$_{-0.24}^{+0.55}$ \\
\noalign{\smallskip}
$\log_{10}(SFR)\ (M_{\odot}/yr)$ & 1.06$_{-0.10}^{+0.12}$  & 0.47$_{-0.10}^{+0.32}$ \\
\noalign{\smallskip}
$sSFR\ (Gyr^{-1})$  & 9.26 $\pm$ 6.02 & 0.76 $\pm$ 0.68 \\
A$_V$ (mag) & 0.19 $\pm$ 0.10 & 0.51 $\pm$ 0.33 \\
$Z_{\ast}$ & 0.006 $\pm$ 0.002 & 0.006 $\pm$ 0.002  \\
Reduced $\chi^2$ & 0.49 & 0.50 \\
\hline
\end{tabular}
\end{table}

\section{Discussion}
\label{sec:discussion}

A broad study on the GRB prompt and afterglow emission, together with its host galaxy is crucial to better constrain the characteristics of GRB\,210610B. The burst prompt emission presents the GRB as a hard burst, as observed by \textit{Swift}, positioning it on the top part for the long GRBs region of the hardness-ratio vs. T$_{90}$ relation \citep[see e.g.][]{Lien2016}. The isotropic equivalent energy, together with the observed peak energy, shows GRB\,210610B to be fully consistent with the ``Amati'' relation \citep{Amati2002AA, Amati2006MNRAS} for long GRBs. The prompt, high-energy emission of GRB\,210610B has been analysed with the Fermi data by \citep{Chen2022}. They find the prompt emission is best fit by a hybrid jet model \citep{GaoZhang2015} in which a hot fireball component dominates emission at the beginning, while a Poynting flux component supersedes at later times. \cite{Chen2022} find this results to be consistent with the magnetar model as a plausible central engine \citep{Metzger2011}.

\subsection{Host galaxy}

The SED analysis of GRB\,210610B host reveals a galaxy with a low stellar mass, consistent with a dwarf galaxy that is actively forming stars and has low extinction.  At the same redshift, the companion is more massive than the GRB host, with a higher stellar mass but lower SFR. The extinction for this galaxy is also higher than the one of the GRB host assuming the same extinction law.

From the afterglow spectrum, we find that GRB\,210610B is embedded in an environment with low amounts of \ion{Fe}{ii} and somewhat higher values for \ion{Mg}{ii} and \ion{Mg}{i}. 
The low amount of \ion{Fe}{} could be indicative of low number of SN\,Ia in the host, since these explosions are the main sources of \ion{Fe}{} \citep{Pagel2009}. This could mean that either there is an intrinsic lack of this type of stellar explosions near the absorber site or that the system itself is too young to have been \ion{Fe}{}-enriched via SN\,Ia. However, the absence of fine-structure lines in the spectrum does not allow us to determine the distance of the absorbing clouds to the explosion site and, therefore, the afterglow spectrum could be tracing gas in the external parts of the host galaxy, where we would expect it to be less enriched. Nevertheless, we find a larger relative value for the EWs of \ion{Mg}{}, although still low compared to the mean value found for GRB sites \citep{deUgartePostigo2012}. \ion{Mg}{} is released into the ISM on the explosion of massive stars \citep{Pagel2009}. Together with the low amount of \ion{Fe}{}, this might suggest that the host galaxy is a very young system. This is also supported by the extinction we measure from the SED fit. The A$_{V}$ value may indicate that the host galaxy has a low amount of dust which is expected for a low metallicity and, therefore, for a young system.

\subsection{Afterglow}

The GRB afterglow follows a decay-plateau-decay behaviour with an initial decay that is well fitted by a broken power-law with an optical spectral slope at $\sim$0.08\ days of $\beta_{opt}=0.43_{-0.05}^{+0.05}$ and a $\beta_{XR}\sim0.83$. Afterwards, the light-curve enters in a plateau phase that last $\sim$0.247 days to finally change to the final decay at $\sim$0.326\ days after burst. This final decay is better fitted by a power-law with $\beta=0.869_{-0.007}^{+0.003}$, consistent within errors with $\beta_{XR}$ at $\sim$0.08\ days after GRB. The change in the spectral slope at $\sim$0.08\ days might be indicating a spectral break at the beginning of the plateau phase. The SED fit shows negligible extinction on the line of sight towards the GRB and a low X-ray Hydrogen column density, as compared to $N_H$ values for relatively low redshift GRBs \citep{Campana2010}. This low $N_H$ together with the low E(B-V), is opposite to what would be expected for low redshift bursts, where a higher dust-to-gas ratio is expected for lowly $N_H$ X-ray absorbed bursts \citep{Campana2010}.

\subsection{Polarisation}

Our polarisation observations match three very important stages of the GRB\,210610B afterglow light-curve. The first polarimetry measurements were performed 0.1205 days after  burst, right after the light-curve enters the plateau. The second  epoch at $\sim$0.26 days is close to the end of this plateau phase, almost consistent with the break of the optical light-curve at 0.326\ days after the GRB. After this, the light curve undergoes its final decay, where a final  polarimetry epoch was  observed. The afterglow shows to be polarised at the beginning and at the end of the light curve evolution but not around the optical break at 0.326 days after burst where the polarisation drops to zero with a small rise to 0.61\% in the next observation, slightly above the P$_{\textnormal{Host ISP}}$ limit but only at a 2.5-$\sigma$ level. In the final decay, the polarisation goes up  to 2.27\%  and then, down to 1.69\% as the afterglow fades away. We find that \textit{b-} and \textit{z-}bands show polarisation values consistent with zero at a close epoch to the 2.5-$\sigma$ \textit{R-}band measurement. Therefore the multi-band observations do not allow us to asses chromaticity/achromaticity on the afterglow polarisation. The PA varies between the 3-$\sigma$ detections $\sim$54$^\circ$. The measured polarisation is well consistent with prior measurements for GRB afterglow linear polarimetry, as shown in Fig. \ref{Fig:comparison_polarisation}. 

One important aspect when determining the intrinsic polarisation is to constrain any contribution from the dust in the host galaxy. A possible polarisation from the MW has been removed during the analysis (see Sect. \ref{sec:polarimetry_analysis}). The SED fit to the GRB light-curve results in a negligible value for the afterglow extinction on the line-of-sight and the inferred upper limit for the GRB host  ISP is rather small compared to errors of the measured 3-$\sigma$ polarisation detections (see Sect. \ref{sec:polarimetry_analysis}). This means that either the polarisation contribution of the host galaxy along the line-of-sight is well below P$_{\textnormal{Host ISP}}$ limit or that this contribution would be cancelling the afterglow polarisation. It is the relatively high values we measure for the 3-$\sigma$ detections and the very low limit for the host galaxy polarisation what lead us to assume the host contribution to be negligible. This low extinction also support the scenario in which we interpret the polarisation as intrinsic to the GRB afterglow.  This is further confirmed by the polarisation non-detection on the second epoch, which is an non-direct measurement of the host ISP.

\subsection{Theoretical Interpretations of the polarisation signals}

The  first polarimetry observations show a rather high linear polarisation degree $\sim 4\%$ at $t\sim 0.1205$ day.  Considering that the observations were carried out during a shallow decay/plateau phase, a non-negligible fraction of optical photons might originate from refreshed shocks in the original ejecta from the central engine. As previous polarimetry studies of the early afterglow indicate that ejecta from the central engine contain large-scale ordered magnetic fields, at least for a subgroup of GRBs if not all \citep[e. g.][]{Mundell2013}, the refreshed shock emission can be polarised due to the ordered magnetic fields in the ejecta. The combination of the polarised refreshed shock emission and unpolarised forward shock emission is likely to give low/intermediate polarisation signals. For GRB 191016A, \cite{Shrestha2022} report the detection of polarisation signals $P\sim 5-15\%$ which are coincident with the start of the plateau phase. An energy injection model has been discussed to explain the coincidence. 

The optical light curve starts to decline at $t\sim 0.2$ days (see the top panel of Fig.\ref{Fig:polarisation_plot}). This indicates that the energy injection stops around that time and the optical band is dominated by the forward shock emission well before the  second polarisation  epoch is conducted at $t\sim 0.24$ day. The magnetic fields in the forward shock region (the shocked ambient medium) are conventionally assumed to be generated locally by microscopic instabilities in shocks \citep[see e.g.][]{Medvedev1999}, and expected to be highly tangled. The PD of the forward shock emission is expected to be zero if the line of sight does not run along the jet edge. The low polarisation at $t\sim 0.24-0.28$ days can be naturally explained if the optical emission is dominated by the forward shock emission. 

Due to the relativistic beaming effect, the observer can see only a small visible region (a small patch  with an angular size of $1/\Gamma$, located around the point at which the line of sight intersects the jet) instead of the entire surface of a shock front.  The visible region appears as a ring due to a relativistic limb-brightening effect \citep{Granot1999}. Synchrotron emission from each small segment of the ring can be polarised if the random magnetic fields parallel and perpendicular to the shock normal have different averaged strengths. However, the net PD is zero because of the symmetry of the visible region.

As the forward shock slows down, the angular size of the visible region grows as $1/\Gamma \propto t^{3/8}$ (ISM) or $t^{1/4}$ (wind medium). Eventually, a part of the ring is located outside the jet edge and the emission region becomes asymmetric. This might have happened by the  third polarisation epoch at $t\sim 1.27$ day. Between the  2$^{nd}$ and 3$^{rd}$ epochs, the angular size of the visible region can grow by a factor of $\sim 1.7$ (ISM) or 1.4 (wind medium). Optical linear polarisation measurements have been carried out for many late GRB afterglows typically several hours to a few days after the prompt gamma-ray emission \citep[e. g.][]{covinogotz16}. This is the period in which a jet break is expected to occur. The detection or upper limits of the linear PD are generally low  (less than a few percent), which possibly indicates that the shock-generated random magnetic fields parallel and perpendicular to the shock normal have similar averaged strengths. The polarisation signals $P\sim 2\%$ at $t\sim 1.27$ days might be explained in this geometrical model. If the large-scale magnetic fields in the eject is toroidal, the PA change between $t\sim 0.24$ days and $t\sim 1.27$ days is expected to be 0 or 90 degree. However, the large-scale magnetic fields in the ejecta can be largely distorted before it injects energy into the forward shock (or the original magnetic structure can be very different from the toroidal configuration). The position angle change can be any value. 
However, this model predicts a steeper decline at late times.  Even in the non-spreading jet model, the expected decay index is $\alpha= 3(p-1)/4 + 3/4 \sim 2.05$ (ISM) or $(3p-1)/4+1/2 \sim 2.30$ (wind medium) for $p=2.74$, which is steeper than the observed value $\alpha =1.85\pm 0.04$. If we rely on the rather high value of $p$ obtained from the SED modelling, we can rule out this geometrical model. 

The nature of magnetic fields generated in shock instabilities are not fully understood yet. The microscopic scale tangled magnetic fields may decay so rapidly in the downstream of the shock that it could not account for the observed synchrotron flux \citep[e. g.][]{Sironi2015}. Alternatively, the forward shock region could have magnetic field turbulence on large scales, comparable to the width of shocked region $\sim R/16\Gamma$ \citep[e. g.][]{Sironi2007}. In this case, the PD and PA temporally change in a random manner, and PD $\sim 70\%/\sqrt{N}$, where $N$ is the number of patches with coherent magnetic field within the angular scale $1/\Gamma$ \cite{GruzinovWaxman1999}. \cite{Kuwata2023} constructed a semi-analytic model of varying large-scale turbulent field in the forward shock region, for which they performed numerical calculations in the case of isotropic turbulence and zero viewing angle and obtained randomly varying PD on a timescale of hours at a level of $\sim 1-3\%$ and PA with changes that are not limited to $90^\circ$. These properties appear consistent with our data of GRB\,210610B. If hydrodynamic-scale turbulent magnetic fields are assumed, we have two possible scenarios 1) the $\sim 4\%, \sim 0.2\%$ and $\sim 2\%$ polarisation signals are all due to turbulent magnetic fields, 2) the  $\sim 4\%$ polarisation signal is due to polarised refreshed shock emission, and the other two are due to turbulent magnetic fields. 

For non-spreading top-hat jets with microscopic scale tangled fields, the PD light curve would have two maxima around a jet break, with the polarisation PA changing by 90 degree between the first and the second maximum \citep{Ghisellini1999, Sari1999}. A possible jet break associated with the PA change of 90 degree has been detected for GRB\,121024A \citep{Wiersema2014}. We have studied a non-axisymmetric top-hat jet model (homogeneous jets with elliptic jet edge) to see whether the main features can be explained in this model. Such jets might be produced due to the interaction between jets and stellar envelop/neutron star wind ejecta (see Fig. 1 in \cite{Lamb2022}) or jet precession \citep{Huang2019}. For non-axisymmetric jets, the PA change can be different from 90 degree. Following \cite{Sari1999}, we have estimated the polarisation light curve and the PA change around a jet break. However, we find that this model does not work for this event. The main reasons are as follows; a) this model also predicts the steep decay at late times as discussed in the geometrical model. b) We need to fine-tune the timing of the  second polarisation measurement (we need to place the observation precisely at the "valley" of the polarisation curve), or equivalently we need to fine-tune a combination of parameters which gives the jet break time. 
c) At the “valley”, the polarization needs to be very small (0.18-0.6\%), compared to the earlier observation of $\sim 4\%$. To achieve this, Stoke’s \textit{u} also needs to be almost zero when \textit{q} flips the sign. According to our rough parameter search  (the geometrical parameters are ellipse eccentricity, ellipse orientation and viewing angle), the eccentricity of the jet edge needs to be smaller than roughly about 0.3. The small eccentricity does not allow the PA change to be significantly different from 90 degrees (in our example case to fit the polarisation light curve, the PA change is about 80 degrees). 

\section{Conclusions}
\label{sec:conclusions}

GRB\,210610B presents an exceptional scenario to perform polarimetry on a GRB optical afterglow. The light-curve follows a decay-plateau-decay trend with a break after the plateau phase to a steep decay of the light-curve. The SED modelling of the afterglow, from X-rays to optical, indicates a dust free line-of-sight towards the GRB as well as a low $N_H$. This negligible amount of dust is confirmed by the low A$_V$ value we derive for the host galaxy  and further confirmed for the polarisation non-detection on the second polarimetry epoch. We also find the GRB is embedded in a low mass galaxy that seems to have a low amount of metals which is indicative of a very young system.

The low amount of dust we find for GRB\,210610B allow us to study the GRB afterglow intrinsic polarisation. The optical afterglow is polarised at the beginning of the plateau phase of the light-curve, disappears around the break achromatically and reappears in the final decay of the light curve. In this complex behaviour, the first epoch seems to be dominated by the refreshed shock, which could explain the high polarisation value, while in the following epoch the polarisation degree drops to zero, as the forward shock would be dominating the optical emission. In the final decay of the light-curve, the polarisation goes up to $\sim 2\%$ which could be explained assuming a geometrical model or hydrodinamics-scale turbulent magnetic fields.

Some models predict GRB afterglow polarisation evolving from high polarisation at early stages, while the prompt emission or the refreshed shocks dominate, followed by a fast decay of polarisation often reaching a zero polarisation. Afterwards, the polarisation increases again to moderate/low values, including changes in the polarisation position angle (see \cite{covinogotz16} for a review). To better understand GRB polarisation evolution we should pursue two approaches: On one hand we need to study polarisation throughout different GRB light curve phases, on the other hand we need to obtain larger samples of GRBs.

\begin{acknowledgements}
      JFAF acknowledges support from the Spanish Ministerio de Ciencia, Innovaci\'on y Universidades through the grant PRE2018-086507. 
       AR acknowledge support from PRIN-MIUR 2017 (grant 20179ZF5KS).
      
      This work is partly based on observations made with the Gran Telescopio Canarias (GTC), installed in the Spanish Observatorio del Roque de los Muchachos of the Instituto de Astrof\' isica de Canarias, in the island of La Palma.

      Partly based on observations collected at the Centro Astron\'omico Hispano en Andaluc\'ia (CAHA) at Calar Alto, operated jointly by Junta de Andaluc\'ia and Consejo Superior de Investigaciones Cient\'ificas (IAA-CSIC).
      
      This work made use of the GRBspec database \url{https://grbspec.eu}. This work has made extensive use of IRAF and Python, particularly with \texttt{astropy} \citep[\url{http://www.astropy.org}]{2013A&A...558A..33A}, \texttt{matplotlib} \citep{matplotlib}, \texttt{photutils} \citep{photutils}, \texttt{numpy} \citep{harris2020array}, and \texttt{Scipy} \citep{Scipy}.
      
\end{acknowledgements}

\bibliographystyle{aa} 
\bibliography{references}

\begin{thebibliography}{120}
\expandafter\ifx\csname natexlab\endcsname\relax\def\natexlab#1{#1}\fi

\bibitem[{{Abbott} {et~al.}(2017){Abbott}, {Abbott}, {Abbott}, {Acernese},
  {Ackley}, {Adams}, {Adams}, {Addesso}, {Adhikari}, {Adya}, {Affeldt},
  {Afrough}, {Agarwal}, {Agathos}, {Agatsuma}, {Aggarwal}, {Aguiar}, {Aiello},
  {Ain}, {Ajith}, {Allen}, {Allen}, {Allocca}, {Aloy}, {Altin}, {Amato},
  {Ananyeva}, {Anderson}, {Anderson}, {Angelova}, {Antier}, {Appert}, {Arai},
  {Araya}, {Areeda}, {Arnaud}, {Arun}, {Ascenzi}, {Ashton}, {Ast}, {Aston},
  {Astone}, {Atallah}, {Aufmuth}, {Aulbert}, {AultONeal}, {Austin},
  {Avila-Alvarez}, {Babak}, {Bacon}, {Bader}, {Bae}, {Baker}, {Baldaccini},
  {Ballardin}, {Ballmer}, {Banagiri}, {Barayoga}, {Barclay}, {Barish},
  {Barker}, {Barkett}, {Barone}, {Barr}, {Barsotti}, {Barsuglia}, {Barta},
  {Bartlett}, {Bartos}, {Bassiri}, {Basti}, {Batch}, {Bawaj}, {Bayley},
  {Bazzan}, {B{\'e}csy}, {Beer}, {Bejger}, {Belahcene}, {Bell}, {Berger},
  {Bergmann}, {Bero}, {Berry}, {Bersanetti}, {Bertolini}, {Betzwieser},
  {Bhagwat}, {Bhandare}, {Bilenko}, {Billingsley}, {Billman}, {Birch},
  {Birney}, {Birnholtz}, {Biscans}, {Biscoveanu}, {Bisht}, {Bitossi}, {Biwer},
  {Bizouard}, {Blackburn}, {Blackman}, {Blair}, {Blair}, {Blair}, {Bloemen},
  {Bock}, {Bode}, {Boer}, {Bogaert}, {Bohe}, {Bondu}, {Bonilla}, {Bonnand},
  {Boom}, {Bork}, {Boschi}, {Bose}, {Bossie}, {Bouffanais}, {Bozzi},
  {Bradaschia}, {Brady}, {Branchesi}, {Brau}, {Briant}, {Brillet}, {Brinkmann},
  {Brisson}, {Brockill}, {Broida}, {Brooks}, {Brown}, {Brown}, {Brunett},
  {Buchanan}, {Buikema}, {Bulik}, {Bulten}, {Buonanno}, {Buskulic}, {Buy},
  {Byer}, {Cabero}, {Cadonati}, {Cagnoli}, {Cahillane}, {Calder{\'o}n
  Bustillo}, {Callister}, {Calloni}, {Camp}, {Canepa}, {Canizares}, {Cannon},
  {Cao}, {Cao}, {Capano}, {Capocasa}, {Carbognani}, {Caride}, {Carney},
  {Casanueva Diaz}, {Casentini}, {Caudill}, {Cavagli{\`a}}, {Cavalier},
  {Cavalieri}, {Cella}, {Cepeda}, {Cerd{\'a}-Dur{\'a}n}, {Cerretani},
  {Cesarini}, {Chamberlin}, {Chan}, {Chao}, {Charlton}, {Chase},
  {Chassande-Mottin}, {Chatterjee}, {Chatziioannou}, {Cheeseboro}, {Chen},
  {Chen}, {Chen}, {Cheng}, {Chia}, {Chincarini}, {Chiummo}, {Chmiel}, {Cho},
  {Cho}, {Chow}, {Christensen}, {Chu}, {Chua}, {Chua}, {Chung}, {Chung},
  {Ciani}, {Ciolfi}, {Cirelli}, {Cirone}, {Clara}, {Clark}, {Clearwater},
  {Cleva}, {Cocchieri}, {Coccia}, {Cohadon}, {Cohen}, {Colla}, {Collette},
  {Cominsky}, {Constancio}, {Conti}, {Cooper}, {Corban}, {Corbitt},
  {Cordero-Carri{\'o}n}, {Corley}, {Cornish}, {Corsi}, {Cortese}, {Costa},
  {Coughlin}, {Coughlin}, {Coulon}, {Countryman}, {Couvares}, {Covas}, {Cowan},
  {Coward}, {Cowart}, {Coyne}, {Coyne}, {Creighton}, {Creighton}, {Cripe},
  {Crowder}, {Cullen}, {Cumming}, {Cunningham}, {Cuoco}, {Dal Canton},
  {D{\'a}lya}, {Danilishin}, {D'Antonio}, {Danzmann}, {Dasgupta}, {Da Silva
  Costa}, {Dattilo}, {Dave}, {Davier}, {Davis}, {Daw}, {Day}, {De}, {DeBra},
  {Degallaix}, {De Laurentis}, {Del{\'e}glise}, {Del Pozzo}, {Demos}, {Denker},
  {Dent}, {De Pietri}, {Dergachev}, {De Rosa}, {DeRosa}, {De Rossi}, {DeSalvo},
  {de Varona}, {Devenson}, {Dhurandhar}, {D{\'\i}az}, {Di Fiore}, {Di
  Giovanni}, {Di Girolamo}, {Di Lieto}, {Di Pace}, {Di Palma}, {Di Renzo},
  {Doctor}, {Dolique}, {Donovan}, {Dooley}, {Doravari}, {Dorrington},
  {Douglas}, {Dovale {\'A}lvarez}, {Downes}, {Drago}, {Dreissigacker},
  {Driggers}, {Du}, {Ducrot}, {Dupej}, {Dwyer}, {Edo}, {Edwards}, {Effler},
  {Eggenstein}, {Ehrens}, {Eichholz}, {Eikenberry}, {Eisenstein}, {Essick},
  {Estevez}, {Etienne}, {Etzel}, {Evans}, {Evans}, {Factourovich}, {Fafone},
  {Fair}, {Fairhurst}, {Fan}, {Farinon}, {Farr}, {Farr}, {Fauchon-Jones},
  {Favata}, {Fays}, {Fee}, {Fehrmann}, {Feicht}, {Fejer}, {Fernandez-Galiana},
  {Ferrante}, {Ferreira}, {Ferrini}, {Fidecaro}, {Finstad}, {Fiori},
  {Fiorucci}, {Fishbach}, {Fisher}, {Fitz-Axen}, {Flaminio}, {Fletcher},
  {Fong}, {Font}, {Forsyth}, {Forsyth}, {Fournier}, {Frasca}, {Frasconi},
  {Frei}, {Freise}, {Frey}, {Frey}, {Fries}, {Fritschel}, {Frolov}, {Fulda},
  {Fyffe}, {Gabbard}, {Gadre}, {Gaebel}, {Gair}, {Gammaitoni}, {Ganija},
  {Gaonkar}, {Garcia-Quiros}, {Garufi}, {Gateley}, {Gaudio}, {Gaur},
  {Gayathri}, {Gehrels}, {Gemme}, {Genin}, {Gennai}, {George}, {George},
  {Gergely}, {Germain}, {Ghonge}, {Ghosh}, {Ghosh}, {Ghosh}, {Giaime},
  {Giardina}, {Giazotto}, {Gill}, {Glover}, {Goetz}, {Goetz}, {Gomes},
  {Goncharov}, {Gonz{\'a}lez}, {Gonzalez Castro}, {Gopakumar}, {Gorodetsky},
  {Gossan}, {Gosselin}, {Gouaty}, {Grado}, {Graef}, {Granata}, {Grant}, {Gras},
  {Gray}, {Greco}, {Green}, {Gretarsson}, {Groot}, {Grote}, {Grunewald},
  {Gruning}, {Guidi}, {Guo}, {Gupta}, {Gupta}, {Gushwa}, {Gustafson},
  {Gustafson}, {Halim}, {Hall}, {Hall}, {Hamilton}, {Hammond}, {Haney},
  {Hanke}, {Hanks}, {Hanna}, {Hannam}, {Hannuksela}, {Hanson}, {Hardwick},
  {Harms}, {Harry}, {Harry}, {Hart}, {Haster}, {Haughian}, {Healy}, {Heidmann},
  {Heintze}, {Heitmann}, {Hello}, {Hemming}, {Hendry}, {Heng}, {Hennig},
  {Heptonstall}, {Heurs}, {Hild}, {Hinderer}, {Hoak}, {Hofman}, {Holt}, {Holz},
  {Hopkins}, {Horst}, {Hough}, {Houston}, {Howell}, {Hreibi}, {Hu}, {Huerta},
  {Huet}, {Hughey}, {Husa}, {Huttner}, {Huynh-Dinh}, {Indik}, {Inta}, {Intini},
  {Isa}, {Isac}, {Isi}, {Iyer}, {Izumi}, {Jacqmin}, {Jani}, {Jaranowski},
  {Jawahar}, {Jim{\'e}nez-Forteza}, {Johnson}, {Johnson-McDaniel}, {Jones},
  {Jones}, {Jonker}, {Ju}, {Junker}, {Kalaghatgi}, {Kalogera}, {Kamai},
  {Kandhasamy}, {Kang}, {Kanner}, {Kapadia}, {Karki}, {Karvinen}, {Kasprzack},
  {Kastaun}, {Katolik}, {Katsavounidis}, {Katzman}, {Kaufer}, {Kawabe},
  {K{\'e}f{\'e}lian}, {Keitel}, {Kemball}, {Kennedy}, {Kent}, {Key}, {Khalili},
  {Khan}, {Khan}, {Khan}, {Khazanov}, {Kijbunchoo}, {Kim}, {Kim}, {Kim}, {Kim},
  {Kim}, {Kim}, {Kimbrell}, {King}, {King}, {Kinley-Hanlon}, {Kirchhoff},
  {Kissel}, {Kleybolte}, {Klimenko}, {Knowles}, {Koch}, {Koehlenbeck}, {Koley},
  {Kondrashov}, {Kontos}, {Korobko}, {Korth}, {Kowalska}, {Kozak},
  {Kr{\"a}mer}, {Kringel}, {Krishnan}, {Kr{\'o}lak}, {Kuehn}, {Kumar}, {Kumar},
  {Kumar}, {Kuo}, {Kutynia}, {Kwang}, {Lackey}, {Lai}, {Landry}, {Lang},
  {Lange}, {Lantz}, {Lanza}, {Lartaux-Vollard}, {Lasky}, {Laxen}, {Lazzarini},
  {Lazzaro}, {Leaci}, {Leavey}, {Lee}, {Lee}, {Lee}, {Lee}, {Lee}, {Lehmann},
  {Lenon}, {Leonardi}, {Leroy}, {Letendre}, {Levin}, {Li}, {Linker},
  {Littenberg}, {Liu}, {Lo}, {Lockerbie}, {London}, {Lord}, {Lorenzini},
  {Loriette}, {Lormand}, {Losurdo}, {Lough}, {Lousto}, {Lovelace}, {L{\"u}ck},
  {Lumaca}, {Lundgren}, {Lynch}, {Ma}, {Macas}, {Macfoy}, {Machenschalk},
  {MacInnis}, {Macleod}, {Maga{\~n}a Hernandez}, {Maga{\~n}a-Sandoval},
  {Maga{\~n}a Zertuche}, {Magee}, {Majorana}, {Maksimovic}, {Man}, {Mandic},
  {Mangano}, {Mansell}, {Manske}, {Mantovani}, {Marchesoni}, {Marion},
  {M{\'a}rka}, {M{\'a}rka}, {Markakis}, {Markosyan}, {Markowitz}, {Maros},
  {Marquina}, {Martelli}, {Martellini}, {Martin}, {Martin}, {Martynov},
  {Mason}, {Massera}, {Masserot}, {Massinger}, {Masso-Reid}, {Mastrogiovanni},
  {Matas}, {Matichard}, {Matone}, {Mavalvala}, {Mazumder}, {McCarthy},
  {McClelland}, {McCormick}, {McCuller}, {McGuire}, {McIntyre}, {McIver},
  {McManus}, {McNeill}, {McRae}, {McWilliams}, {Meacher}, {Meadors}, {Mehmet},
  {Meidam}, {Mejuto-Villa}, {Melatos}, {Mendell}, {Mercer}, {Merilh},
  {Merzougui}, {Meshkov}, {Messenger}, {Messick}, {Metzdorff}, {Meyers},
  {Miao}, {Michel}, {Middleton}, {Mikhailov}, {Milano}, {Miller}, {Miller},
  {Miller}, {Millhouse}, {Milovich-Goff}, {Minazzoli}, {Minenkov}, {Ming},
  {Mishra}, {Mitra}, {Mitrofanov}, {Mitselmakher}, {Mittleman}, {Moffa},
  {Moggi}, {Mogushi}, {Mohan}, {Mohapatra}, {Montani}, {Moore}, {Moraru},
  {Moreno}, {Morriss}, {Mours}, {Mow-Lowry}, {Mueller}, {Muir}, {Mukherjee},
  {Mukherjee}, {Mukherjee}, {Mukund}, {Mullavey}, {Munch}, {Mu{\~n}iz},
  {Muratore}, {Murray}, {Napier}, {Nardecchia}, {Naticchioni}, {Nayak},
  {Neilson}, {Nelemans}, {Nelson}, {Nery}, {Neunzert}, {Nevin}, {Newport},
  {Newton}, {Ng}, {Nguyen}, {Nichols}, {Nielsen}, {Nissanke}, {Nitz}, {Noack},
  {Nocera}, {Nolting}, {North}, {Nuttall}, {Oberling}, {O'Dea}, {Ogin}, {Oh},
  {Oh}, {Ohme}, {Okada}, {Oliver}, {Oppermann}, {Oram}, {O'Reilly}, {Ormiston},
  {Ortega}, {O'Shaughnessy}, {Ossokine}, {Ottaway}, {Overmier}, {Owen}, {Pace},
  {Page}, {Page}, {Pai}, {Pai}, {Palamos}, {Palashov}, {Palomba}, {Pal-Singh},
  {Pan}, {Pan}, {Pang}, {Pang}, {Pankow}, {Pannarale}, {Pant}, {Paoletti},
  {Paoli}, {Papa}, {Parida}, {Parker}, {Pascucci}, {Pasqualetti},
  {Passaquieti}, {Passuello}, {Patil}, {Patricelli}, {Pearlstone}, {Pedraza},
  {Pedurand}, {Pekowsky}, {Pele}, {Penn}, {Perez}, {Perreca}, {Perri},
  {Pfeiffer}, {Phelps}, {Piccinni}, {Pichot}, {Piergiovanni}, {Pierro},
  {Pillant}, {Pinard}, {Pinto}, {Pirello}, {Pitkin}, {Poe}, {Poggiani},
  {Popolizio}, {Porter}, {Post}, {Powell}, {Prasad}, {Pratt}, {Pratten},
  {Predoi}, {Prestegard}, {Prijatelj}, {Principe}, {Privitera}, {Prodi},
  {Prokhorov}, {Puncken}, {Punturo}, {Puppo}, {P{\"u}rrer}, {Qi}, {Quetschke},
  {Quintero}, {Quitzow-James}, {Raab}, {Rabeling}, {Radkins}, {Raffai}, {Raja},
  {Rajan}, {Rajbhandari}, {Rakhmanov}, {Ramirez}, {Ramos-Buades}, {Rapagnani},
  {Raymond}, {Razzano}, {Read}, {Regimbau}, {Rei}, {Reid}, {Reitze}, {Ren},
  {Reyes}, {Ricci}, {Ricker}, {Rieger}, {Riles}, {Rizzo}, {Robertson}, {Robie},
  {Robinet}, {Rocchi}, {Rolland}, {Rollins}, {Roma}, {Romano}, {Romel},
  {Romie}, {Rosi{\'n}ska}, {Ross}, {Rowan}, {R{\"u}diger}, {Ruggi}, {Rutins},
  {Ryan}, {Sachdev}, {Sadecki}, {Sadeghian}, {Sakellariadou}, {Salconi},
  {Saleem}, {Salemi}, {Samajdar}, {Sammut}, {Sampson}, {Sanchez}, {Sanchez},
  {Sanchis-Gual}, {Sandberg}, {Sanders}, {Sassolas}, {Sathyaprakash},
  {Saulson}, {Sauter}, {Savage}, {Sawadsky}, {Schale}, {Scheel}, {Scheuer},
  {Schmidt}, {Schmidt}, {Schnabel}, {Schofield}, {Sch{\"o}nbeck}, {Schreiber},
  {Schuette}, {Schulte}, {Schutz}, {Schwalbe}, {Scott}, {Scott}, {Seidel},
  {Sellers}, {Sengupta}, {Sentenac}, {Sequino}, {Sergeev}, {Shaddock},
  {Shaffer}, {Shah}, {Shahriar}, {Shaner}, {Shao}, {Shapiro}, {Shawhan},
  {Sheperd}, {Shoemaker}, {Shoemaker}, {Siellez}, {Siemens}, {Sieniawska},
  {Sigg}, {Silva}, {Singer}, {Singh}, {Singhal}, {Sintes}, {Slagmolen},
  {Smith}, {Smith}, {Smith}, {Somala}, {Son}, {Sonnenberg}, {Sorazu},
  {Sorrentino}, {Souradeep}, {Spencer}, {Srivastava}, {Staats}, {Staley},
  {Steinke}, {Steinlechner}, {Steinlechner}, {Steinmeyer}, {Stevenson},
  {Stone}, {Stops}, {Strain}, {Stratta}, {Strigin}, {Strunk}, {Sturani},
  {Stuver}, {Summerscales}, {Sun}, {Sunil}, {Suresh}, {Sutton}, {Swinkels},
  {Szczepa{\'n}czyk}, {Tacca}, {Tait}, {Talbot}, {Talukder}, {Tanner},
  {T{\'a}pai}, {Taracchini}, {Tasson}, {Taylor}, {Taylor}, {Tewari}, {Theeg},
  {Thies}, {Thomas}, {Thomas}, {Thomas}, {Thorne}, {Thorne}, {Thrane},
  {Tiwari}, {Tiwari}, {Tokmakov}, {Toland}, {Tonelli}, {Tornasi},
  {Torres-Forn{\'e}}, {Torrie}, {T{\"o}yr{\"a}}, {Travasso}, {Traylor},
  {Trinastic}, {Tringali}, {Trozzo}, {Tsang}, {Tse}, {Tso}, {Tsukada}, {Tsuna},
  {Tuyenbayev}, {Ueno}, {Ugolini}, {Unnikrishnan}, {Urban}, {Usman},
  {Vahlbruch}, {Vajente}, {Valdes}, {van Bakel}, {van Beuzekom}, {van den
  Brand}, {Van Den Broeck}, {Vander-Hyde}, {van der Schaaf}, {van Heijningen},
  {van Veggel}, {Vardaro}, {Varma}, {Vass}, {Vas{\'u}th}, {Vecchio},
  {Vedovato}, {Veitch}, {Veitch}, {Venkateswara}, {Venugopalan}, {Verkindt},
  {Vetrano}, {Vicer{\'e}}, {Viets}, {Vinciguerra}, {Vine}, {Vinet}, {Vitale},
  {Vo}, {Vocca}, {Vorvick}, {Vyatchanin}, {Wade}, {Wade}, {Wade}, {Walet},
  {Walker}, {Wallace}, {Walsh}, {Wang}, {Wang}, {Wang}, {Wang}, {Wang}, {Ward},
  {Warner}, {Was}, {Watchi}, {Weaver}, {Wei}, {Weinert}, {Weinstein}, {Weiss},
  {Wen}, {Wessel}, {We{\ss}els}, {Westerweck}, {Westphal}, {Wette}, {Whelan},
  {Whitcomb}, {Whiting}, {Whittle}, {Wilken}, {Williams}, {Williams},
  {Williamson}, {Willis}, {Willke}, {Wimmer}, {Winkler}, {Wipf}, {Wittel},
  {Woan}, {Woehler}, {Wofford}, {Wong}, {Worden}, {Wright}, {Wu}, {Wysocki},
  {Xiao}, {Yamamoto}, {Yancey}, {Yang}, {Yap}, {Yazback}, {Yu}, {Yu}, {Yvert},
  {Zadro{\.z}ny}, {Zanolin}, {Zelenova}, {Zendri}, {Zevin}, {Zhang}, {Zhang},
  {Zhang}, {Zhang}, {Zhao}, {Zhou}, {Zhou}, {Zhu}, {Zhu}, {Zimmerman},
  {Zucker}, {Zweizig}, {(LIGO Scientific Collaboration}, {Virgo Collaboration},
  {Burns}, {Veres}, {Kocevski}, {Racusin}, {Goldstein}, {Connaughton},
  {Briggs}, {Blackburn}, {Hamburg}, {Hui}, {von Kienlin}, {McEnery}, {Preece},
  {Wilson-Hodge}, {Bissaldi}, {Cleveland}, {Gibby}, {Giles}, {Kippen},
  {McBreen}, {Meegan}, {Paciesas}, {Poolakkil}, {Roberts}, {Stanbro},
  {Gamma-ray Burst Monitor}, {Savchenko}, {Ferrigno}, {Kuulkers}, {Bazzano},
  {Bozzo}, {Brandt}, {Chenevez}, {Courvoisier}, {Diehl}, {Domingo}, {Hanlon},
  {Jourdain}, {Laurent}, {Lebrun}, {Lutovinov}, {Mereghetti}, {Natalucci},
  {Rodi}, {Roques}, {Sunyaev}, {Ubertini}, \& {(INTEGRAL}}]{Abbott2017}
{Abbott}, B.~P., {Abbott}, R., {Abbott}, T.~D., {et~al.} 2017, \apjl, 848, L13

\bibitem[{{Amati}(2006)}]{Amati2006MNRAS}
{Amati}, L. 2006, \mnras, 372, 233

\bibitem[{{Amati} {et~al.}(2002){Amati}, {Frontera}, {Tavani}, {in't Zand},
  {Antonelli}, {Costa}, {Feroci}, {Guidorzi}, {Heise}, {Masetti}, {Montanari},
  {Nicastro}, {Palazzi}, {Pian}, {Piro}, \& {Soffitta}}]{Amati2002AA}
{Amati}, L., {Frontera}, F., {Tavani}, M., {et~al.} 2002, \aap, 390, 81

\bibitem[{{Anderson}(2015)}]{FORS2manual}
{Anderson}, J. 2015, FORS2 User Manual, Vol. 96.0 (European Southern
  Observatory)

\bibitem[{{Arnaud}(1996)}]{Arnaud1996}
{Arnaud}, K.~A. 1996, in Astronomical Society of the Pacific Conference Series,
  Vol. 101, Astronomical Data Analysis Software and Systems V, ed. G.~H.
  {Jacoby} \& J.~{Barnes}, 17

\bibitem[{{Astropy Collaboration} {et~al.}(2013){Astropy Collaboration},
  {Robitaille}, {Tollerud}, {Greenfield}, {Droettboom}, {Bray}, {Aldcroft},
  {Davis}, {Ginsburg}, {Price-Whelan}, {Kerzendorf}, {Conley}, {Crighton},
  {Barbary}, {Muna}, {Ferguson}, {Grollier}, {Parikh}, {Nair}, {Unther},
  {Deil}, {Woillez}, {Conseil}, {Kramer}, {Turner}, {Singer}, {Fox}, {Weaver},
  {Zabalza}, {Edwards}, {Azalee Bostroem}, {Burke}, {Casey}, {Crawford},
  {Dencheva}, {Ely}, {Jenness}, {Labrie}, {Lim}, {Pierfederici}, {Pontzen},
  {Ptak}, {Refsdal}, {Servillat}, \& {Streicher}}]{2013A&A...558A..33A}
{Astropy Collaboration}, {Robitaille}, T.~P., {Tollerud}, E.~J., {et~al.} 2013,
  \aap, 558, A33

\bibitem[{{Bagnulo} {et~al.}(2009){Bagnulo}, {Landolfi}, {Landstreet}, {Landi
  Degl'Innocenti}, {Fossati}, \& {Sterzik}}]{Bagnulo2009}
{Bagnulo}, S., {Landolfi}, M., {Landstreet}, J.~D., {et~al.} 2009, \pasp, 121,
  993

\bibitem[{{Barth} {et~al.}(2003){Barth}, {Sari}, {Cohen}, {Goodrich}, {Price},
  {Fox}, {Bloom}, {Soderberg}, \& {Kulkarni}}]{Barth2003}
{Barth}, A.~J., {Sari}, R., {Cohen}, M.~H., {et~al.} 2003, \apjl, 584, L47

\bibitem[{{Barthelmy} {et~al.}(2005){Barthelmy}, {Barbier}, {Cummings},
  {Fenimore}, {Gehrels}, {Hullinger}, {Krimm}, {Markwardt}, {Palmer},
  {Parsons}, {Sato}, {Suzuki}, {Takahashi}, {Tashiro}, \&
  {Tueller}}]{Barthelmy2005SwiftBAT}
{Barthelmy}, S.~D., {Barbier}, L.~M., {Cummings}, J.~R., {et~al.} 2005, \ssr,
  120, 143

\bibitem[{{Bersier} {et~al.}(2003){Bersier}, {McLeod}, {Garnavich}, {Holman},
  {Grav}, {Quinn}, {Kaluzny}, {Challis}, {Bower}, {Wilman}, {Heyl}, {Holland},
  {Hradecky}, {Jha}, \& {Stanek}}]{Bersier2003}
{Bersier}, D., {McLeod}, B., {Garnavich}, P.~M., {et~al.} 2003, \apjl, 583, L63

\bibitem[{{Bertin}(2010)}]{Bertin2010}
{Bertin}, E. 2010, {SWarp: Resampling and Co-adding FITS Images Together},
  Astrophysics Source Code Library, record ascl:1010.068

\bibitem[{{Beuermann} {et~al.}(1999){Beuermann}, {Hessman}, {Reinsch},
  {Nicklas}, {Vreeswijk}, {Galama}, {Rol}, {van Paradijs}, {Kouveliotou},
  {Frontera}, {Masetti}, {Palazzi}, \& {Pian}}]{Beuermann1999a}
{Beuermann}, K., {Hessman}, F.~V., {Reinsch}, K., {et~al.} 1999, \aap, 352, L26

\bibitem[{{Bla{\v{z}}ek} {et~al.}(2020){Bla{\v{z}}ek}, {de Ugarte Postigo},
  {Kann}, {Th{\"o}ne}, {Ag{\"u}{\'\i} Fern{\'a}ndez}, \&
  {Izzo}}]{2020SPIE11452E..18B}
{Bla{\v{z}}ek}, M., {de Ugarte Postigo}, A., {Kann}, D.~A., {et~al.} 2020, in
  Society of Photo-Optical Instrumentation Engineers (SPIE) Conference Series,
  Vol. 11452, Society of Photo-Optical Instrumentation Engineers (SPIE)
  Conference Series, 1145218

\bibitem[{{Boquien} {et~al.}(2019){Boquien}, {Burgarella}, {Roehlly}, {Buat},
  {Ciesla}, {Corre}, {Inoue}, \& {Salas}}]{CIGALE2019}
{Boquien}, M., {Burgarella}, D., {Roehlly}, Y., {et~al.} 2019, \aap, 622, A103

\bibitem[{Bradley {et~al.}(2019)Bradley, Sipőcz, Robitaille, Tollerud,
  Vinícius, Deil, Barbary, Wilson, Busko, Günther, Cara, Conseil, Droettboom,
  Bostroem, Bray, Bratholm, Lim, Craig, Barentsen, Pascual, Donath, Greco,
  Perren, Kerzendorf, de~Val-Borro, Dencheva, de~Albernaz~Ferreira, Souchereau,
  D'Eugenio, \& Weaver}]{photutils}
Bradley, L., Sipőcz, B., Robitaille, T., {et~al.} 2019, astropy/photutils:
  v0.7.2

\bibitem[{{Brivio} {et~al.}(2022){Brivio}, {Covino}, {D'Avanzo}, {Wiersema},
  {Maund}, {Bernardini}, {Campana}, \& {Melandri}}]{Brivio2022}
{Brivio}, R., {Covino}, S., {D'Avanzo}, P., {et~al.} 2022, \aap, 666, A179

\bibitem[{{Bruzual} \& {Charlot}(2003)}]{BC03}
{Bruzual}, G. \& {Charlot}, S. 2003, \mnras, 344, 1000

\bibitem[{{Burgarella} {et~al.}(2005){Burgarella}, {Buat}, \&
  {Iglesias-P{\'a}ramo}}]{CIGALE2005}
{Burgarella}, D., {Buat}, V., \& {Iglesias-P{\'a}ramo}, J. 2005, \mnras, 360,
  1413

\bibitem[{{Burns} {et~al.}(2023){Burns}, {Svinkin}, {Fenimore}, {Kann},
  {Ag{\"u}{\'\i} Fern{\'a}ndez}, {Frederiks}, {Hamburg}, {Lesage}, {Temiraev},
  {Tsvetkova}, {Bissaldi}, {Briggs}, {Dalessi}, {Dunwoody}, {Fletcher},
  {Goldstein}, {Hui}, {Hristov}, {Kocevski}, {Lysenko}, {Mailyan}, {Mangan},
  {McBreen}, {Racusin}, {Ridnaia}, {Roberts}, {Ulanov}, {Veres},
  {Wilson-Hodge}, \& {Wood}}]{Burns2023}
{Burns}, E., {Svinkin}, D., {Fenimore}, E., {et~al.} 2023, \apjl, 946, L31

\bibitem[{{Burrows} {et~al.}(2005){Burrows}, {Hill}, {Nousek}, {Kennea},
  {Wells}, {Osborne}, {Abbey}, {Beardmore}, {Mukerjee}, {Short}, {Chincarini},
  {Campana}, {Citterio}, {Moretti}, {Pagani}, {Tagliaferri}, {Giommi},
  {Capalbi}, {Tamburelli}, {Angelini}, {Cusumano}, {Br{\"a}uninger}, {Burkert},
  \& {Hartner}}]{Burrows2005SwiftXRT}
{Burrows}, D.~N., {Hill}, J.~E., {Nousek}, J.~A., {et~al.} 2005, \ssr, 120, 165

\bibitem[{{Calzetti} {et~al.}(2000){Calzetti}, {Armus}, {Bohlin}, {Kinney},
  {Koornneef}, \& {Storchi-Bergmann}}]{Calzetti2000}
{Calzetti}, D., {Armus}, L., {Bohlin}, R.~C., {et~al.} 2000, \apj, 533, 682

\bibitem[{{Campana} {et~al.}(2010){Campana}, {Th{\"o}ne}, {de Ugarte Postigo},
  {Tagliaferri}, {Moretti}, \& {Covino}}]{Campana2010}
{Campana}, S., {Th{\"o}ne}, C.~C., {de Ugarte Postigo}, A., {et~al.} 2010,
  \mnras, 402, 2429

\bibitem[{Caswell {et~al.}(2020)Caswell, Droettboom, Lee, Hunter, Firing,
  de~Andrade, Hoffmann, Stansby, Klymak, Varoquaux, Nielsen, Root, Elson, May,
  Dale, Lee, Seppänen, McDougall, Straw, Hobson, Gohlke, Yu, Ma, Vincent,
  Silvester, Moad, Kniazev, hannah, \& Ernest}]{matplotlib}
Caswell, T.~A., Droettboom, M., Lee, A., {et~al.} 2020, matplotlib/matplotlib:
  REL: v3.2.2

\bibitem[{{Cepa} {et~al.}(2000){Cepa}, {Aguiar}, {Escalera},
  {Gonzalez-Serrano}, {Joven-Alvarez}, {Peraza}, {Rasilla}, {Rodriguez-Ramos},
  {Gonzalez}, {Cobos Duenas}, {Sanchez}, {Tejada}, {Bland-Hawthorn},
  {Militello}, \& {Rosa}}]{OSIRIS2000SPIE}
{Cepa}, J., {Aguiar}, M., {Escalera}, V.~G., {et~al.} 2000, in Society of
  Photo-Optical Instrumentation Engineers (SPIE) Conference Series, Vol. 4008,
  Optical and IR Telescope Instrumentation and Detectors, ed. M.~{Iye} \& A.~F.
  {Moorwood}, 623--631

\bibitem[{{Chabrier}(2003)}]{Chabrier2003}
{Chabrier}, G. 2003, \pasp, 115, 763

\bibitem[{{Chen} {et~al.}(2022){Chen}, {Peng}, {Du}, \& {Yin}}]{Chen2022}
{Chen}, J.-M., {Peng}, Z.-Y., {Du}, T.-T., \& {Yin}, Y. 2022, \apj, 932, 25

\bibitem[{{Christensen} {et~al.}(2011){Christensen}, {Fynbo}, {Prochaska},
  {Th{\"o}ne}, {de Ugarte Postigo}, \& {Jakobsson}}]{Christensen2011}
{Christensen}, L., {Fynbo}, J.~P.~U., {Prochaska}, J.~X., {et~al.} 2011, \apj,
  727, 73

\bibitem[{{Cikota} {et~al.}(2017){Cikota}, {Patat}, {Cikota}, \&
  {Faran}}]{Cikota2017}
{Cikota}, A., {Patat}, F., {Cikota}, S., \& {Faran}, T. 2017, \mnras, 464, 4146

\bibitem[{{Covino} \& {Gotz}(2016)}]{covinogotz16}
{Covino}, S. \& {Gotz}, D. 2016, Astronomical and Astrophysical Transactions,
  29, 205

\bibitem[{{Covino} {et~al.}(1999){Covino}, {Lazzati}, {Ghisellini}, {Saracco},
  {Campana}, {Chincarini}, {di Serego}, {Cimatti}, {Vanzi}, {Pasquini},
  {Haardt}, {Israel}, {Stella}, \& {Vietri}}]{Covino1999}
{Covino}, S., {Lazzati}, D., {Ghisellini}, G., {et~al.} 1999, \aap, 348, L1

\bibitem[{{Covino} {et~al.}(2003){Covino}, {Malesani}, {Ghisellini}, {Lazzati},
  {di Serego Alighieri}, {Stefanon}, {Cimatti}, {Della Valle}, {Fiore},
  {Goldoni}, {Kawai}, {Israel}, {Le Floc'h}, {Mirabel}, {Ricker}, {Saracco},
  {Stella}, {Tagliaferri}, \& {Zerbi}}]{Covino2003}
{Covino}, S., {Malesani}, D., {Ghisellini}, G., {et~al.} 2003, \aap, 400, L9

\bibitem[{{Dale} {et~al.}(2014){Dale}, {Helou}, {Magdis}, {Armus},
  {D{\'\i}az-Santos}, \& {Shi}}]{dale14}
{Dale}, D.~A., {Helou}, G., {Magdis}, G.~E., {et~al.} 2014, \apj, 784, 83

\bibitem[{{de Ugarte Postigo} {et~al.}(2014){de Ugarte Postigo}, {Blazek},
  {Janout}, {Sprimont}, {Th{\"o}ne}, {Gorosabel}, \&
  {S{\'a}nchez-Ram{\'\i}rez}}]{2014SPIE.9152E..0BD}
{de Ugarte Postigo}, A., {Blazek}, M., {Janout}, P., {et~al.} 2014, in Society
  of Photo-Optical Instrumentation Engineers (SPIE) Conference Series, Vol.
  9152, Software and Cyberinfrastructure for Astronomy III, 91520B

\bibitem[{{de Ugarte Postigo} {et~al.}(2005){de Ugarte Postigo},
  {Castro-Tirado}, {Gorosabel}, {J{\'o}hannesson}, {Bj{\"o}rnsson},
  {Gudmundsson}, {Bremer}, {Pak}, {Tanvir}, {Castro Cer{\'o}n}, {Guzyi},
  {Jel{\'\i}nek}, {Klose}, {P{\'e}rez-Ram{\'\i}rez}, {Aceituno}, {Campo
  Bagat{\'\i}n}, {Covino}, {Cardiel}, {Fathkullin}, {Henden}, {Huferath},
  {Kurata}, {Malesani}, {Mannucci}, {Ruiz-Lapuente}, {Sokolov}, {Thiele},
  {Wisotzki}, {Antonelli}, {Bartolini}, {Boattini}, {Guarnieri}, {Piccioni},
  {Pizzichini}, {del Principe}, {di Paola}, {Fugazza}, {Ghisellini}, {Hunt},
  {Konstantinova}, {Masetti}, {Palazzi}, {Pian}, {Stefanon}, {Testa}, \&
  {Tristram}}]{deUgartePostigo2005}
{de Ugarte Postigo}, A., {Castro-Tirado}, A.~J., {Gorosabel}, J., {et~al.}
  2005, \aap, 443, 841

\bibitem[{{de Ugarte Postigo} {et~al.}(2012){de Ugarte Postigo}, {Fynbo},
  {Th{\"o}ne}, {Christensen}, {Gorosabel}, {Milvang-Jensen}, {Schulze},
  {Jakobsson}, {Wiersema}, {S{\'a}nchez-Ram{\'\i}rez}, {Leloudas}, {Zafar},
  {Malesani}, \& {Hjorth}}]{deUgartePostigo2012}
{de Ugarte Postigo}, A., {Fynbo}, J.~P.~U., {Th{\"o}ne}, C.~C., {et~al.} 2012,
  \aap, 548, A11

\bibitem[{{Dhillon} {et~al.}(2021){Dhillon}, {Bezawada}, {Black}, {Dixon},
  {Gamble}, {Gao}, {Henry}, {Kerry}, {Littlefair}, {Lunney}, {Marsh}, {Miller},
  {Parsons}, {Ashley}, {Breedt}, {Brown}, {Dyer}, {Green}, {Pelisoli},
  {Sahman}, {Wild}, {Ives}, {Mehrgan}, {Stegmeier}, {Dubbeldam}, {Morris},
  {Osborn}, {Wilson}, {Casares}, {Mu{\~n}oz-Darias}, {Pall{\'e}},
  {Rodr{\'\i}guez-Gil}, {Shahbaz}, {Torres}, {de Ugarte Postigo},
  {Cabrera-Lavers}, {Corradi}, {Dom{\'\i}nguez}, \&
  {Garc{\'\i}a-Alvarez}}]{Dhillon2021}
{Dhillon}, V.~S., {Bezawada}, N., {Black}, M., {et~al.} 2021, \mnras, 507, 350

\bibitem[{{Evans} {et~al.}(2009){Evans}, {Beardmore}, {Page}, {Osborne},
  {O'Brien}, {Willingale}, {Starling}, {Burrows}, {Godet}, {Vetere}, {Racusin},
  {Goad}, {Wiersema}, {Angelini}, {Capalbi}, {Chincarini}, {Gehrels}, {Kennea},
  {Margutti}, {Morris}, {Mountford}, {Pagani}, {Perri}, {Romano}, \&
  {Tanvir}}]{Evans2009a}
{Evans}, P.~A., {Beardmore}, A.~P., {Page}, K.~L., {et~al.} 2009, \mnras, 397,
  1177

\bibitem[{{Evans} {et~al.}(2007){Evans}, {Beardmore}, {Page}, {Tyler},
  {Osborne}, {Goad}, {O'Brien}, {Vetere}, {Racusin}, {Morris}, {Burrows},
  {Capalbi}, {Perri}, {Gehrels}, \& {Romano}}]{Evans2007a}
{Evans}, P.~A., {Beardmore}, A.~P., {Page}, K.~L., {et~al.} 2007, \aap, 469,
  379

\bibitem[{{Frederiks} {et~al.}(2021){Frederiks}, {Golenetskii}, {Lysenko},
  {Ridnaia}, {Svinkin}, {Tsvetkova}, {Ulanov}, {Cline}, \& {Konus-Wind
  Team}}]{FrederiksGCNKonusWind}
{Frederiks}, D., {Golenetskii}, S., {Lysenko}, A., {et~al.} 2021, GRB
  Coordinates Network, 30196, 1

\bibitem[{{Fynbo} {et~al.}(2021){Fynbo}, {Izzo}, {de Ugarte Postigo},
  {Malesani}, \& {Pursimo}}]{Fynbo2021}
{Fynbo}, J.~P.~U., {Izzo}, L., {de Ugarte Postigo}, A., {Malesani}, D.~B., \&
  {Pursimo}, T. 2021, GRB Coordinates Network, 30182, 1

\bibitem[{{Galama} {et~al.}(1998){Galama}, {Vreeswijk}, {van Paradijs},
  {Kouveliotou}, {Augusteijn}, {B{\"o}hnhardt}, {Brewer}, {Doublier},
  {Gonzalez}, {Leibundgut}, {Lidman}, {Hainaut}, {Patat}, {Heise}, {in't Zand},
  {Hurley}, {Groot}, {Strom}, {Mazzali}, {Iwamoto}, {Nomoto}, {Umeda},
  {Nakamura}, {Young}, {Suzuki}, {Shigeyama}, {Koshut}, {Kippen}, {Robinson},
  {de Wildt}, {Wijers}, {Tanvir}, {Greiner}, {Pian}, {Palazzi}, {Frontera},
  {Masetti}, {Nicastro}, {Feroci}, {Costa}, {Piro}, {Peterson}, {Tinney},
  {Boyle}, {Cannon}, {Stathakis}, {Sadler}, {Begam}, \& {Ianna}}]{Galama1998}
{Galama}, T.~J., {Vreeswijk}, P.~M., {van Paradijs}, J., {et~al.} 1998, \nat,
  395, 670

\bibitem[{{Gao} \& {Zhang}(2015)}]{GaoZhang2015}
{Gao}, H. \& {Zhang}, B. 2015, \apj, 801, 103

\bibitem[{{Garz{\'o}n} {et~al.}(2022){Garz{\'o}n}, {Balcells}, {Gallego},
  {Gry}, {Guzm{\'a}n}, {Hammersley}, {Herrero}, {Mu{\~n}oz-Tu{\~n}{\'o}n},
  {Pell{\'o}}, {Prieto}, {Bourrec}, {Cabello}, {Cardiel},
  {Gonz{\'a}lez-Fern{\'a}ndez}, {Laporte}, {Milliard}, {Pascual}, {Patrick},
  {Patr{\'o}n}, {Ram{\'\i}rez-Alegr{\'\i}a}, \& {Streblyanska}}]{Garzon2022}
{Garz{\'o}n}, F., {Balcells}, M., {Gallego}, J., {et~al.} 2022, \aap, 667, A107

\bibitem[{{Ghisellini} \& {Lazzati}(1999)}]{Ghisellini1999}
{Ghisellini}, G. \& {Lazzati}, D. 1999, \mnras, 309, L7

\bibitem[{{Gill} {et~al.}(2021){Gill}, {Kole}, \& {Granot}}]{Gill2021}
{Gill}, R., {Kole}, M., \& {Granot}, J. 2021, Galaxies, 9, 82

\bibitem[{{Gompertz} {et~al.}(2022){Gompertz}, {Ravasio}, {Nicholl}, {Levan},
  {Metzger}, {Oates}, {Lamb}, {Fong}, {Malesani}, {Rastinejad}, {Tanvir},
  {Evans}, {Jonker}, {Page}, \& {Pe'er}}]{Gompertz2022}
{Gompertz}, B.~P., {Ravasio}, M.~E., {Nicholl}, M., {et~al.} 2022, Nature
  Astronomy [\eprint[arXiv]{2205.05008}]

\bibitem[{{Gonz{\'a}lez-Gait{\'a}n} {et~al.}(2020){Gonz{\'a}lez-Gait{\'a}n},
  {Mour{\~a}o}, {Patat}, {Anderson}, {Cikota}, {Wiersema}, {Higgins}, \&
  {Silva}}]{Santiago2020}
{Gonz{\'a}lez-Gait{\'a}n}, S., {Mour{\~a}o}, A.~M., {Patat}, F., {et~al.} 2020,
  \aap, 634, A70

\bibitem[{{Gorosabel} {et~al.}(2010){Gorosabel}, {de Ugarte Postigo},
  {Castro-Tirado}, {Agudo}, {Jel{\'\i}nek}, {Leon}, {Augusteijn}, {Fynbo},
  {Hjorth}, {Micha{\l}owski}, {Xu}, {Ferrero}, {Kann}, {Klose}, {Rossi},
  {Madrid}, {Llorente}, {Bremer}, \& {Winters}}]{Gorosabel2010}
{Gorosabel}, J., {de Ugarte Postigo}, A., {Castro-Tirado}, A.~J., {et~al.}
  2010, \aap, 522, A14

\bibitem[{{Gorosabel} {et~al.}(2004){Gorosabel}, {Rol}, {Covino},
  {Castro-Tirado}, {Castro Cer{\'o}n}, {Lazzati}, {Hjorth}, {Malesani}, {Della
  Valle}, {di Serego Alighieri}, {Fiore}, {Fruchter}, {Fynbo}, {Ghisellini},
  {Goldoni}, {Greiner}, {Israel}, {Kaper}, {Kawai}, {Klose}, {Kouveliotou}, {Le
  Floc'h}, {Masetti}, {Mirabel}, {M{\"o}ller}, {Ortolani}, {Palazzi}, {Pian},
  {Rhoads}, {Ricker}, {Saracco}, {Stella}, {Tagliaferri}, {Tanvir}, {van den
  Heuvel}, {Vietri}, {Vreeswijk}, {Wijers}, \& {Zerbi}}]{Gorosabel2004}
{Gorosabel}, J., {Rol}, E., {Covino}, S., {et~al.} 2004, \aap, 422, 113

\bibitem[{{Granot} {et~al.}(1999){Granot}, {Piran}, \& {Sari}}]{Granot1999}
{Granot}, J., {Piran}, T., \& {Sari}, R. 1999, \apj, 513, 679

\bibitem[{{Greiner} {et~al.}(2003){Greiner}, {Klose}, {Reinsch}, {Martin
  Schmid}, {Sari}, {Hartmann}, {Kouveliotou}, {Rau}, {Palazzi}, {Straubmeier},
  {Stecklum}, {Zharikov}, {Tovmassian}, {B{\"a}rnbantner}, {Ries}, {Jehin},
  {Henden}, {Kaas}, {Grav}, {Hjorth}, {Pedersen}, {Wijers}, {Kaufer}, {Park},
  {Williams}, \& {Reimer}}]{greiner2003}
{Greiner}, J., {Klose}, S., {Reinsch}, K., {et~al.} 2003, \nat, 426, 157

\bibitem[{{Gruzinov} \& {Waxman}(1999)}]{GruzinovWaxman1999}
{Gruzinov}, A. \& {Waxman}, E. 1999, \apj, 511, 852

\bibitem[{Harris {et~al.}(2020)Harris, Millman, van~der Walt, Gommers,
  Virtanen, Cournapeau, Wieser, Taylor, Berg, Smith, Kern, Picus, Hoyer, van
  Kerkwijk, Brett, Haldane, del R{\'{i}}o, Wiebe, Peterson,
  G{\'{e}}rard-Marchant, Sheppard, Reddy, Weckesser, Abbasi, Gohlke, \&
  Oliphant}]{harris2020array}
Harris, C.~R., Millman, K.~J., van~der Walt, S.~J., {et~al.} 2020, Nature, 585,
  357

\bibitem[{{Hjorth} {et~al.}(2003){Hjorth}, {Sollerman}, {M{\o}ller}, {Fynbo},
  {Woosley}, {Kouveliotou}, {Tanvir}, {Greiner}, {Andersen}, {Castro-Tirado},
  {Castro Cer{\'o}n}, {Fruchter}, {Gorosabel}, {Jakobsson}, {Kaper}, {Klose},
  {Masetti}, {Pedersen}, {Pedersen}, {Pian}, {Palazzi}, {Rhoads}, {Rol}, {van
  den Heuvel}, {Vreeswijk}, {Watson}, \& {Wijers}}]{Hjorth2003}
{Hjorth}, J., {Sollerman}, J., {M{\o}ller}, P., {et~al.} 2003, \nat, 423, 847

\bibitem[{{Ho} {et~al.}(2022){Ho}, {Perley}, {Yao}, {Svinkin}, {de Ugarte
  Postigo}, {Perley}, {Kann}, {Burns}, {Andreoni}, {Bellm}, {Bissaldi},
  {Bloom}, {Brink}, {Dekany}, {Drake}, {Ag{\"u}{\'\i} Fern{\'a}ndez},
  {Filippenko}, {Frederiks}, {Graham}, {Hristov}, {Kasliwal}, {Kulkarni},
  {Kumar}, {Laher}, {Lysenko}, {Mailyan}, {Malacaria}, {Miller}, {Poolakkil},
  {Riddle}, {Ridnaia}, {Rusholme}, {Savchenko}, {Sollerman}, {Th{\"o}ne},
  {Tsvetkova}, {Ulanov}, \& {von Kienlin}}]{Ho2022}
{Ho}, A. Y.~Q., {Perley}, D.~A., {Yao}, Y., {et~al.} 2022, \apj, 938, 85

\bibitem[{{Huang} {et~al.}(2019){Huang}, {Lin}, {Liu}, {Ren}, {Wang}, {Liu}, \&
  {Liang}}]{Huang2019}
{Huang}, B.-Q., {Lin}, D.-B., {Liu}, T., {et~al.} 2019, \mnras, 487, 3214

\bibitem[{{Jakobsson} {et~al.}(2012){Jakobsson}, {Hjorth}, {Malesani},
  {Chapman}, {Fynbo}, {Tanvir}, {Milvang-Jensen}, {Vreeswijk}, {Letawe}, \&
  {Starling}}]{Jakobsson2012}
{Jakobsson}, P., {Hjorth}, J., {Malesani}, D., {et~al.} 2012, \apj, 752, 62

\bibitem[{{Kobayashi} {et~al.}(1997){Kobayashi}, {Piran}, \&
  {Sari}}]{Kobayashi1997}
{Kobayashi}, S., {Piran}, T., \& {Sari}, R. 1997, \apj, 490, 92

\bibitem[{{Kouveliotou} {et~al.}(1993){Kouveliotou}, {Meegan}, {Fishman},
  {Bhat}, {Briggs}, {Koshut}, {Paciesas}, \& {Pendleton}}]{Kouveliotou1993}
{Kouveliotou}, C., {Meegan}, C.~A., {Fishman}, G.~J., {et~al.} 1993, \apjl,
  413, L101

\bibitem[{{Krimm} {et~al.}(2021){Krimm}, {Barthelmy}, {Cummings}, {Laha},
  {Lien}, {Markwardt}, {Page}, {Palmer}, {Sakamoto}, {Stamatikos}, \&
  {Ukwatta}}]{KrimmGCNRefinedBAT}
{Krimm}, H.~A., {Barthelmy}, S.~D., {Cummings}, J.~R., {et~al.} 2021, GRB
  Coordinates Network, 30207, 1

\bibitem[{{Kuwata} {et~al.}(2023){Kuwata}, {Toma}, {Kimura}, {Tomita}, \&
  {Shimoda}}]{Kuwata2023}
{Kuwata}, A., {Toma}, K., {Kimura}, S.~S., {Tomita}, S., \& {Shimoda}, J. 2023,
  \apj, 943, 118

\bibitem[{{Lamb} {et~al.}(2022){Lamb}, {Nativi}, {Rosswog}, {Kann}, {Levan},
  {Lundman}, \& {Tanvir}}]{Lamb2022}
{Lamb}, G.~P., {Nativi}, L., {Rosswog}, S., {et~al.} 2022, Universe, 8, 612

\bibitem[{{Lazzati} {et~al.}(2003){Lazzati}, {Covino}, {di Serego Alighieri},
  {Ghisellini}, {Vernet}, {Le Floc'h}, {Fugazza}, {Di Tomaso}, {Malesani},
  {Masetti}, {Pian}, {Oliva}, \& {Stella}}]{Lazzati2003}
{Lazzati}, D., {Covino}, S., {di Serego Alighieri}, S., {et~al.} 2003, \aap,
  410, 823

\bibitem[{{Levan} {et~al.}(2023{\natexlab{a}}){Levan}, {Gompertz}, {Salafia},
  {Bulla}, {Burns}, {Hotokezaka}, {Izzo}, {Lamb}, {Malesani}, {Oates},
  {Ravasio}, {Rouco Escorial}, {Schneider}, {Sarin}, {Schulze}, {Tanvir},
  {Ackley}, {Anderson}, {Brammer}, {Christensen}, {Dhillon}, {Evans},
  {Fausnaugh}, {Fong}, {Fruchter}, {Fryer}, {Fynbo}, {Gaspari}, {Heintz},
  {Hjorth}, {Kennea}, {Kennedy}, {Laskar}, {Leloudas}, {Mandel},
  {Martin-Carrillo}, {Metzger}, {Nicholl}, {Nugent}, {Palmerio}, {Pugliese},
  {Rastinejad}, {Rhodes}, {Rossi}, {Smartt}, {Stevance}, {Tohuvavohu}, {van der
  Horst}, {Vergani}, {Watson}, {Barclay}, {Bhirombhakdi}, {Breedt}, {Breeveld},
  {Brown}, {Campana}, {Chrimes}, {D'Avanzo}, {D'Elia}, {De Pasquale}, {Dyer},
  {Galloway}, {Garbutt}, {Green}, {Hartmann}, {Jakobsson}, {Kerry},
  {Langeroodi}, {Leung}, {Littlefair}, {Munday}, {O'Brien}, {Parsons},
  {Pelisoli}, {Saccardi}, {Sahman}, {Salvaterra}, {Sbarufatti}, {Steeghs},
  {Tagliaferri}, {Th{\"o}ne}, {de Ugarte Postigo}, \& {Kann}}]{Levan2023b}
{Levan}, A., {Gompertz}, B.~P., {Salafia}, O.~S., {et~al.} 2023{\natexlab{a}},
  arXiv e-prints, arXiv:2307.02098

\bibitem[{{Levan} {et~al.}(2023{\natexlab{b}}){Levan}, {Malesani}, {Gompertz},
  {Nugent}, {Nicholl}, {Oates}, {Perley}, {Rastinejad}, {Metzger}, {Schulze},
  {Stanway}, {Inkenhaag}, {Zafar}, {Ag{\"u}{\'\i} Fern{\'a}ndez}, {Chrimes},
  {Bhirombhakdi}, {de Ugarte Postigo}, {Fong}, {Fruchter}, {Fragione}, {Fynbo},
  {Gaspari}, {Heintz}, {Hjorth}, {Jakobsson}, {Jonker}, {Lamb}, {Mandel},
  {Mandhai}, {Ravasio}, {Sollerman}, \& {Tanvir}}]{Levan2023a}
{Levan}, A.~J., {Malesani}, D.~B., {Gompertz}, B.~P., {et~al.}
  2023{\natexlab{b}}, Nature Astronomy, 7, 976

\bibitem[{{Lien} {et~al.}(2016){Lien}, {Sakamoto}, {Barthelmy}, {Baumgartner},
  {Cannizzo}, {Chen}, {Collins}, {Cummings}, {Gehrels}, {Krimm}, {Markwardt},
  {Palmer}, {Stamatikos}, {Troja}, \& {Ukwatta}}]{Lien2016}
{Lien}, A., {Sakamoto}, T., {Barthelmy}, S.~D., {et~al.} 2016, \apj, 829, 7

\bibitem[{{Lipkin} {et~al.}(2004){Lipkin}, {Ofek}, {Gal-Yam}, {Leibowitz},
  {Poznanski}, {Kaspi}, {Polishook}, {Kulkarni}, {Fox}, {Berger}, {Mirabal},
  {Halpern}, {Bureau}, {Fathi}, {Price}, {Peterson}, {Frebel}, {Schmidt},
  {Orosz}, {Fitzgerald}, {Bloom}, {van Dokkum}, {Bailyn}, {Buxton}, \&
  {Barsony}}]{Lipkin2004}
{Lipkin}, Y.~M., {Ofek}, E.~O., {Gal-Yam}, A., {et~al.} 2004, \apj, 606, 381

\bibitem[{{Magalhaes} {et~al.}(2003){Magalhaes}, {Pereyra}, {Dominici}, \&
  {Abraham}}]{magalhaes2003}
{Magalhaes}, A.~M., {Pereyra}, A., {Dominici}, T., \& {Abraham}, Z. 2003, GRB
  Coordinates Network, 2163, 1

\bibitem[{{Malacaria} {et~al.}(2021){Malacaria}, {Hristov}, \& {Fermi GBM
  Team}}]{MalacariaGCNFermiGBM}
{Malacaria}, C., {Hristov}, B., \& {Fermi GBM Team}. 2021, GRB Coordinates
  Network, 30199, 1

\bibitem[{{Mandarakas} {et~al.}(2023){Mandarakas}, {Blinov}, {Aguilera-Dena},
  {Romanopoulos}, {Pavlidou}, {Tassis}, {Antoniadis}, {Kiehlmann}, {Lychoudis},
  \& {Tsemperof Kataivatis}}]{Mandarakas2023}
{Mandarakas}, N., {Blinov}, D., {Aguilera-Dena}, D.~R., {et~al.} 2023, \aap,
  670, A144

\bibitem[{{Masetti} {et~al.}(2003){Masetti}, {Palazzi}, {Pian}, {Simoncelli},
  {Hunt}, {Maiorano}, {Levan}, {Christensen}, {Rol}, {Savaglio}, {Falomo},
  {Castro-Tirado}, {Hjorth}, {Delsanti}, {Pannella}, {Mohan}, {Pandey},
  {Sagar}, {Amati}, {Burud}, {Castro Cer{\'o}n}, {Frontera}, {Fruchter},
  {Fynbo}, {Gorosabel}, {Kaper}, {Klose}, {Kouveliotou}, {Nicastro},
  {Pedersen}, {Rhoads}, {Salamanca}, {Tanvir}, {Vreeswijk}, {Wijers}, \& {van
  den Heuvel}}]{Masetti2003}
{Masetti}, N., {Palazzi}, E., {Pian}, E., {et~al.} 2003, \aap, 404, 465

\bibitem[{{Medvedev} \& {Loeb}(1999)}]{Medvedev1999}
{Medvedev}, M.~V. \& {Loeb}, A. 1999, \apj, 526, 697

\bibitem[{{M{\'e}sz{\'a}ros} \& {Rees}(1997)}]{Meszaros1997}
{M{\'e}sz{\'a}ros}, P. \& {Rees}, M.~J. 1997, \apj, 476, 232

\bibitem[{{Metzger} {et~al.}(2011){Metzger}, {Giannios}, {Thompson},
  {Bucciantini}, \& {Quataert}}]{Metzger2011}
{Metzger}, B.~D., {Giannios}, D., {Thompson}, T.~A., {Bucciantini}, N., \&
  {Quataert}, E. 2011, \mnras, 413, 2031

\bibitem[{{Morgan} {et~al.}(2008){Morgan}, {Vanden Berk}, {Roming}, {Nousek},
  {Koch}, {Breeveld}, {de Pasquale}, {Holland}, {Kuin}, {Page}, \&
  {Still}}]{Morgan2008}
{Morgan}, A.~N., {Vanden Berk}, D.~E., {Roming}, P.~W.~A., {et~al.} 2008, \apj,
  683, 913

\bibitem[{{Mundell} {et~al.}(2013){Mundell}, {Kopa{\v{c}}}, {Arnold}, {Steele},
  {Gomboc}, {Kobayashi}, {Harrison}, {Smith}, {Guidorzi}, {Virgili},
  {Melandri}, \& {Japelj}}]{Mundell2013}
{Mundell}, C.~G., {Kopa{\v{c}}}, D., {Arnold}, D.~M., {et~al.} 2013, \nat, 504,
  119

\bibitem[{{Nagao} {et~al.}(2022){Nagao}, {Patat}, {Maeda}, {Baade}, {Mattila},
  {Taubenberger}, {Kotak}, {Cikota}, {Kuncarayakti}, {Bulla}, \&
  {Maund}}]{Nagao2022}
{Nagao}, T., {Patat}, F., {Maeda}, K., {et~al.} 2022, \apjl, 941, L4

\bibitem[{{Negro} {et~al.}(2023){Negro}, {Di Lalla}, {Omodei}, {Veres},
  {Silvestri}, {Manfreda}, {Burns}, {Baldini}, {Costa}, {Ehlert}, {Kennea},
  {Liodakis}, {Marshall}, {Mereghetti}, {Middei}, {Muleri}, {O'Dell},
  {Roberts}, {Romani}, {Sgr{\'o}}, {Terashima}, {Tiengo}, {Viscolo}, {Di
  Marco}, {La Monaca}, {Latronico}, {Matt}, {Perri}, {Puccetti}, {Poutanen},
  {Ratheesh}, {Rogantini}, {Slane}, {Soffitta}, {Lindfors}, {Nilsson},
  {Kasikov}, {Marscher}, {Tavecchio}, {Cibrario}, {Gunji}, {Malacaria},
  {Paggi}, {Yang}, {Zane}, {Weisskopf}, {Agudo}, {Antonelli}, {Bachetti},
  {Baumgartner}, {Bellazzini}, {Bianchi}, {Bongiorno}, {Bonino}, {Brez},
  {Bucciantini}, {Capitanio}, {Castellano}, {Cavazzuti}, {Chen}, {Ciprini}, {De
  Rosa}, {Del Monte}, {Di Gesu}, {Donnarumma}, {Doroshenko}, {Dovc̆iak},
  {Enoto}, {Evangelista}, {Fabiani}, {Ferrazzoli}, {Garcia}, {Hayashida},
  {Heyl}, {Iwakiri}, {Jorstad}, {Kaaret}, {Karas}, {Kislat}, {Kitaguchi},
  {Kolodziejczak}, {Krawczynski}, {Maldera}, {Marin}, {Marinucci}, {Mitsuishi},
  {Mizuno}, {Ng}, {Oppedisano}, {Papitto}, {Pavlov}, {Peirson},
  {Pesce-Rollins}, {Petrucci}, {Pilia}, {Possenti}, {Ramsey}, {Rankin},
  {Spandre}, {Swartz}, {Tamagawa}, {Taverna}, {Tawara}, {Tennant}, {Thomas},
  {Tombesi}, {Trois}, {Tsygankov}, {Turolla}, {Vink}, {Wu}, \&
  {Xie}}]{Negro2023}
{Negro}, M., {Di Lalla}, N., {Omodei}, N., {et~al.} 2023, \apjl, 946, L21

\bibitem[{{Noll} {et~al.}(2009){Noll}, {Burgarella}, {Giovannoli}, {Buat},
  {Marcillac}, \& {Mu{\~n}oz-Mateos}}]{CIGALE2009}
{Noll}, S., {Burgarella}, D., {Giovannoli}, E., {et~al.} 2009, \aap, 507, 1793

\bibitem[{{Oke}(1974)}]{Oke1974}
{Oke}, J.~B. 1974, \apjs, 27, 21

\bibitem[{{Page} {et~al.}(2021){Page}, {Gropp}, {Kennea}, {Marshall}, {Palmer},
  {Siegel}, \& {Neil Gehrels Swift Observatory Team}}]{Page2021BATDetection}
{Page}, K.~L., {Gropp}, J.~D., {Kennea}, J.~A., {et~al.} 2021, GRB Coordinates
  Network, 30170, 1

\bibitem[{{Pagel}(2009)}]{Pagel2009}
{Pagel}, B. E.~J. 2009, {Nucleosynthesis and Chemical Evolution of Galaxies}

\bibitem[{{Patat} \& {Romaniello}(2006)}]{Patat2006}
{Patat}, F. \& {Romaniello}, M. 2006, \pasp, 118, 146

\bibitem[{{Patat} \& {Taubenberger}(2011)}]{Patat2011CAFOS}
{Patat}, F. \& {Taubenberger}, S. 2011, \aap, 529, A57

\bibitem[{{Pei}(1992)}]{Pei1992a}
{Pei}, Y.~C. 1992, \apj, 395, 130

\bibitem[{{Perley}(2021)}]{Perley30216}
{Perley}, D.~A. 2021, GRB Coordinates Network, 30216, 1

\bibitem[{{Piran}(1999)}]{Piran1999}
{Piran}, T. 1999, \physrep, 314, 575

\bibitem[{{Planck Collaboration} {et~al.}(2014){Planck Collaboration}, {Ade},
  {Aghanim}, {Armitage-Caplan}, {Arnaud}, {Ashdown}, {Atrio-Barandela},
  {Aumont}, {Baccigalupi}, {Banday}, {Barreiro}, {Bartlett}, {Battaner},
  {Benabed}, {Beno{\^\i}t}, {Benoit-L{\'e}vy}, {Bernard}, {Bersanelli},
  {Bielewicz}, {Bobin}, {Bock}, {Bonaldi}, {Bond}, {Borrill}, {Bouchet},
  {Bridges}, {Bucher}, {Burigana}, {Butler}, {Calabrese}, {Cappellini},
  {Cardoso}, {Catalano}, {Challinor}, {Chamballu}, {Chary}, {Chen}, {Chiang},
  {Chiang}, {Christensen}, {Church}, {Clements}, {Colombi}, {Colombo},
  {Couchot}, {Coulais}, {Crill}, {Curto}, {Cuttaia}, {Danese}, {Davies},
  {Davis}, {de Bernardis}, {de Rosa}, {de Zotti}, {Delabrouille}, {Delouis},
  {D{\'e}sert}, {Dickinson}, {Diego}, {Dolag}, {Dole}, {Donzelli}, {Dor{\'e}},
  {Douspis}, {Dunkley}, {Dupac}, {Efstathiou}, {Elsner}, {En{\ss}lin},
  {Eriksen}, {Finelli}, {Forni}, {Frailis}, {Fraisse}, {Franceschi}, {Gaier},
  {Galeotta}, {Galli}, {Ganga}, {Giard}, {Giardino}, {Giraud-H{\'e}raud},
  {Gjerl{\o}w}, {Gonz{\'a}lez-Nuevo}, {G{\'o}rski}, {Gratton}, {Gregorio},
  {Gruppuso}, {Gudmundsson}, {Haissinski}, {Hamann}, {Hansen}, {Hanson},
  {Harrison}, {Henrot-Versill{\'e}}, {Hern{\'a}ndez-Monteagudo}, {Herranz},
  {Hildebrandt}, {Hivon}, {Hobson}, {Holmes}, {Hornstrup}, {Hou}, {Hovest},
  {Huffenberger}, {Jaffe}, {Jaffe}, {Jewell}, {Jones}, {Juvela},
  {Keih{\"a}nen}, {Keskitalo}, {Kisner}, {Kneissl}, {Knoche}, {Knox}, {Kunz},
  {Kurki-Suonio}, {Lagache}, {L{\"a}hteenm{\"a}ki}, {Lamarre}, {Lasenby},
  {Lattanzi}, {Laureijs}, {Lawrence}, {Leach}, {Leahy}, {Leonardi},
  {Le{\'o}n-Tavares}, {Lesgourgues}, {Lewis}, {Liguori}, {Lilje},
  {Linden-V{\o}rnle}, {L{\'o}pez-Caniego}, {Lubin}, {Mac{\'\i}as-P{\'e}rez},
  {Maffei}, {Maino}, {Mandolesi}, {Maris}, {Marshall}, {Martin},
  {Mart{\'\i}nez-Gonz{\'a}lez}, {Masi}, {Massardi}, {Matarrese}, {Matthai},
  {Mazzotta}, {Meinhold}, {Melchiorri}, {Melin}, {Mendes}, {Menegoni},
  {Mennella}, {Migliaccio}, {Millea}, {Mitra}, {Miville-Desch{\^e}nes},
  {Moneti}, {Montier}, {Morgante}, {Mortlock}, {Moss}, {Munshi}, {Murphy},
  {Naselsky}, {Nati}, {Natoli}, {Netterfield}, {N{\o}rgaard-Nielsen},
  {Noviello}, {Novikov}, {Novikov}, {O'Dwyer}, {Osborne}, {Oxborrow}, {Paci},
  {Pagano}, {Pajot}, {Paladini}, {Paoletti}, {Partridge}, {Pasian},
  {Patanchon}, {Pearson}, {Pearson}, {Peiris}, {Perdereau}, {Perotto},
  {Perrotta}, {Pettorino}, {Piacentini}, {Piat}, {Pierpaoli}, {Pietrobon},
  {Plaszczynski}, {Platania}, {Pointecouteau}, {Polenta}, {Ponthieu}, {Popa},
  {Poutanen}, {Pratt}, {Pr{\'e}zeau}, {Prunet}, {Puget}, {Rachen}, {Reach},
  {Rebolo}, {Reinecke}, {Remazeilles}, {Renault}, {Ricciardi}, {Riller},
  {Ristorcelli}, {Rocha}, {Rosset}, {Roudier}, {Rowan-Robinson},
  {Rubi{\~n}o-Mart{\'\i}n}, {Rusholme}, {Sandri}, {Santos}, {Savelainen},
  {Savini}, {Scott}, {Seiffert}, {Shellard}, {Spencer}, {Starck}, {Stolyarov},
  {Stompor}, {Sudiwala}, {Sunyaev}, {Sureau}, {Sutton}, {Suur-Uski}, {Sygnet},
  {Tauber}, {Tavagnacco}, {Terenzi}, {Toffolatti}, {Tomasi}, {Tristram},
  {Tucci}, {Tuovinen}, {T{\"u}rler}, {Umana}, {Valenziano}, {Valiviita}, {Van
  Tent}, {Vielva}, {Villa}, {Vittorio}, {Wade}, {Wandelt}, {Wehus}, {White},
  {White}, {Wilkinson}, {Yvon}, {Zacchei}, \& {Zonca}}]{Planck2014}
{Planck Collaboration}, {Ade}, P.~A.~R., {Aghanim}, N., {et~al.} 2014, \aap,
  571, A16

\bibitem[{{Plaszczynski} {et~al.}(2014){Plaszczynski}, {Montier}, {Levrier}, \&
  {Tristram}}]{Plaszczynski2014}
{Plaszczynski}, S., {Montier}, L., {Levrier}, F., \& {Tristram}, M. 2014,
  \mnras, 439, 4048

\bibitem[{{Rastinejad} {et~al.}(2022){Rastinejad}, {Gompertz}, {Levan}, {Fong},
  {Nicholl}, {Lamb}, {Malesani}, {Nugent}, {Oates}, {Tanvir}, {de Ugarte
  Postigo}, {Kilpatrick}, {Moore}, {Metzger}, {Ravasio}, {Rossi}, {Schroeder},
  {Jencson}, {Sand}, {Smith}, {Ag{\"u}{\'\i} Fern{\'a}ndez}, {Berger},
  {Blanchard}, {Chornock}, {Cobb}, {De Pasquale}, {Fynbo}, {Izzo}, {Kann},
  {Laskar}, {Marini}, {Paterson}, {Escorial}, {Sears}, \&
  {Th{\"o}ne}}]{Rastinejad2022}
{Rastinejad}, J.~C., {Gompertz}, B.~P., {Levan}, A.~J., {et~al.} 2022, \nat,
  612, 223

\bibitem[{{Rees} \& {Meszaros}(1994)}]{Rees1994}
{Rees}, M.~J. \& {Meszaros}, P. 1994, \apjl, 430, L93

\bibitem[{{Rol} {et~al.}(2003){Rol}, {Wijers}, {Fynbo}, {Hjorth}, {Gorosabel},
  {Egholm}, {Castro Cer{\'o}n}, {Castro-Tirado}, {Kaper}, {Masetti}, {Palazzi},
  {Pian}, {Tanvir}, {Vreeswijk}, {Kouveliotou}, {M{\o}ller}, {Pedersen},
  {Fruchter}, {Rhoads}, {Burud}, {Salamanca}, \& {Van den Heuvel}}]{Rol2003}
{Rol}, E., {Wijers}, R.~A.~M.~J., {Fynbo}, J.~P.~U., {et~al.} 2003, \aap, 405,
  L23

\bibitem[{{Rol} {et~al.}(2000){Rol}, {Wijers}, {Vreeswijk}, {Kaper}, {Galama},
  {van Paradijs}, {Kouveliotou}, {Masetti}, {Pian}, {Palazzi}, {Frontera}, \&
  {van den Heuvel}}]{Rol2000}
{Rol}, E., {Wijers}, R.~A.~M.~J., {Vreeswijk}, P.~M., {et~al.} 2000, \apj, 544,
  707

\bibitem[{{Roming} {et~al.}(2005){Roming}, {Kennedy}, {Mason}, {Nousek}, {Ahr},
  {Bingham}, {Broos}, {Carter}, {Hancock}, {Huckle}, {Hunsberger}, {Kawakami},
  {Killough}, {Koch}, {McLelland}, {Smith}, {Smith}, {Soto}, {Boyd},
  {Breeveld}, {Holland}, {Ivanushkina}, {Pryzby}, {Still}, \&
  {Stock}}]{Roming2005SwiftUVOT}
{Roming}, P. W.~A., {Kennedy}, T.~E., {Mason}, K.~O., {et~al.} 2005, \ssr, 120,
  95

\bibitem[{{Sari}(1999)}]{Sari1999}
{Sari}, R. 1999, \apjl, 524, L43

\bibitem[{{Sari} \& {Piran}(1997)}]{Sari1997}
{Sari}, R. \& {Piran}, T. 1997, \mnras, 287, 110

\bibitem[{{Sari} {et~al.}(1998){Sari}, {Piran}, \& {Narayan}}]{Sari1998a}
{Sari}, R., {Piran}, T., \& {Narayan}, R. 1998, \apjl, 497, L17

\bibitem[{{Schlafly} \& {Finkbeiner}(2011)}]{Schlafly11}
{Schlafly}, E.~F. \& {Finkbeiner}, D.~P. 2011, \apj, 737, 103

\bibitem[{{Schmidt} {et~al.}(1992){Schmidt}, {Elston}, \&
  {Lupie}}]{Schmidt1992}
{Schmidt}, G.~D., {Elston}, R., \& {Lupie}, O.~L. 1992, \aj, 104, 1563

\bibitem[{{Science Software Branch at STScI}(2012)}]{pyraf}
{Science Software Branch at STScI}. 2012, {PyRAF: Python alternative for IRAF},
  Astrophysics Source Code Library, record ascl:1207.011

\bibitem[{{Serkowski} {et~al.}(1975){Serkowski}, {Mathewson}, \&
  {Ford}}]{Serkowski1975}
{Serkowski}, K., {Mathewson}, D.~S., \& {Ford}, V.~L. 1975, \apj, 196, 261

\bibitem[{{Shrestha} {et~al.}(2022){Shrestha}, {Steele}, {Kobayashi},
  {Jordana-Mitjans}, {Smith}, {Jermak}, {Arnold}, {Mundell}, {Gomboc}, \&
  {Guidorzi}}]{Shrestha2022}
{Shrestha}, M., {Steele}, I.~A., {Kobayashi}, S., {et~al.} 2022, \mnras, 509,
  5964

\bibitem[{{Siegel} {et~al.}(2021){Siegel}, {Baer}, {Page}, \& {Swift/UVOT
  Team}}]{SiegelGCNUVOT}
{Siegel}, M.~H., {Baer}, M., {Page}, K.~L., \& {Swift/UVOT Team}. 2021, GRB
  Coordinates Network, 30247, 1

\bibitem[{{Sironi} \& {Goodman}(2007)}]{Sironi2007}
{Sironi}, L. \& {Goodman}, J. 2007, \apj, 671, 1858

\bibitem[{{Sironi} {et~al.}(2015){Sironi}, {Keshet}, \& {Lemoine}}]{Sironi2015}
{Sironi}, L., {Keshet}, U., \& {Lemoine}, M. 2015, \ssr, 191, 519

\bibitem[{{Steele} {et~al.}(2009){Steele}, {Mundell}, {Smith}, {Kobayashi}, \&
  {Guidorzi}}]{Steele2009}
{Steele}, I.~A., {Mundell}, C.~G., {Smith}, R.~J., {Kobayashi}, S., \&
  {Guidorzi}, C. 2009, \nat, 462, 767

\bibitem[{{Troja} {et~al.}(2022){Troja}, {Fryer}, {O'Connor}, {Ryan},
  {Dichiara}, {Kumar}, {Ito}, {Gupta}, {Wollaeger}, {Norris}, {Kawai},
  {Butler}, {Aryan}, {Misra}, {Hosokawa}, {Murata}, {Niwano}, {Pandey},
  {Kutyrev}, {van Eerten}, {Chase}, {Hu}, {Caballero-Garcia}, \&
  {Castro-Tirado}}]{Troja2022}
{Troja}, E., {Fryer}, C.~L., {O'Connor}, B., {et~al.} 2022, \nat, 612, 228

\bibitem[{{Turnshek} {et~al.}(1990){Turnshek}, {Bohlin}, {Williamson}, {Lupie},
  {Koornneef}, \& {Morgan}}]{Turnshek1990}
{Turnshek}, D.~A., {Bohlin}, R.~C., {Williamson}, R.~L., I., {et~al.} 1990,
  \aj, 99, 1243

\bibitem[{{Uehara} {et~al.}(2012){Uehara}, {Toma}, {Kawabata}, {Chiyonobu},
  {Fukazawa}, {Ikejiri}, {Inoue}, {Itoh}, {Komatsu}, {Miyamoto}, {Mizuno},
  {Nagae}, {Nakaya}, {Ohsugi}, {Sakimoto}, {Sasada}, {Tanaka}, {Uemura},
  {Yamanaka}, {Yamashita}, {Yamazaki}, \& {Yoshida}}]{Uehara2012}
{Uehara}, T., {Toma}, K., {Kawabata}, K.~S., {et~al.} 2012, \apjl, 752, L6

\bibitem[{{Urata} {et~al.}(2023){Urata}, {Toma}, {Covino}, {Wiersema}, {Huang},
  {Shimoda}, {Kuwata}, {Nagao}, {Asada}, {Nagai}, {Takahashi}, {Chung},
  {Petitpas}, {Yamaoka}, {Izzo}, {Fynbo}, {de Ugarte Postigo}, {Arabsalmani},
  \& {Tashiro}}]{Urata2023}
{Urata}, Y., {Toma}, K., {Covino}, S., {et~al.} 2023, Nature Astronomy, 7, 80

\bibitem[{{van Dokkum}(2001)}]{VanDokkum2001}
{van Dokkum}, P.~G. 2001, \pasp, 113, 1420

\bibitem[{Virtanen {et~al.}(2020)Virtanen, Gommers, Oliphant, Haberland, Reddy,
  Cournapeau, Burovski, Peterson, Weckesser, Bright, {van der Walt}, Brett,
  Wilson, Millman, Mayorov, Nelson, Jones, Kern, Larson, Carey, Polat, Feng,
  Moore, {VanderPlas}, Laxalde, Perktold, Cimrman, Henriksen, Quintero, Harris,
  Archibald, Ribeiro, Pedregosa, {van Mulbregt}, \& {SciPy 1.0
  Contributors}}]{Scipy}
Virtanen, P., Gommers, R., Oliphant, T.~E., {et~al.} 2020, Nature Methods, 17,
  261

\bibitem[{{von Kienlin} {et~al.}(2020){von Kienlin}, {Meegan}, {Paciesas},
  {Bhat}, {Bissaldi}, {Briggs}, {Burns}, {Cleveland}, {Gibby}, {Giles},
  {Goldstein}, {Hamburg}, {Hui}, {Kocevski}, {Mailyan}, {Malacaria},
  {Poolakkil}, {Preece}, {Roberts}, {Veres}, \&
  {Wilson-Hodge}}]{vonKienlin2020FermiOnlineCat}
{von Kienlin}, A., {Meegan}, C.~A., {Paciesas}, W.~S., {et~al.} 2020, \apj,
  893, 46

\bibitem[{{Wiersema} {et~al.}(2014){Wiersema}, {Covino}, {Toma}, {van der
  Horst}, {Varela}, {Min}, {Greiner}, {Starling}, {Tanvir}, {Wijers},
  {Campana}, {Curran}, {Fan}, {Fynbo}, {Gorosabel}, {Gomboc}, {G{\"o}tz},
  {Hjorth}, {Jin}, {Kobayashi}, {Kouveliotou}, {Mundell}, {O'Brien}, {Pian},
  {Rowlinson}, {Russell}, {Salvaterra}, {di Serego Alighieri}, {Tagliaferri},
  {Vergani}, {Elliott}, {Fari{\~n}a}, {Hartoog}, {Karjalainen}, {Klose},
  {Knust}, {Levan}, {Schady}, {Sudilovsky}, \& {Willingale}}]{Wiersema2014}
{Wiersema}, K., {Covino}, S., {Toma}, K., {et~al.} 2014, \nat, 509, 201

\bibitem[{{Wiersema} {et~al.}(2012){Wiersema}, {Curran}, {Kr{\"u}hler},
  {Melandri}, {Rol}, {Starling}, {Tanvir}, {van der Horst}, {Covino}, {Fynbo},
  {Goldoni}, {Gorosabel}, {Hjorth}, {Klose}, {Mundell}, {O'Brien}, {Palazzi},
  {Wijers}, {D'Elia}, {Evans}, {Filgas}, {Gomboc}, {Greiner}, {Guidorzi},
  {Kaper}, {Kobayashi}, {Kouveliotou}, {Levan}, {Rossi}, {Rowlinson}, {Steele},
  {de Ugarte Postigo}, \& {Vergani}}]{Wiersema2012}
{Wiersema}, K., {Curran}, P.~A., {Kr{\"u}hler}, T., {et~al.} 2012, \mnras, 426,
  2

\bibitem[{{Wijers} {et~al.}(1999){Wijers}, {Vreeswijk}, {Galama}, {Rol}, {van
  Paradijs}, {Kouveliotou}, {Giblin}, {Masetti}, {Palazzi}, {Pian}, {Frontera},
  {Nicastro}, {Falomo}, {Soffitta}, \& {Piro}}]{Wijers1999}
{Wijers}, R.~A.~M.~J., {Vreeswijk}, P.~M., {Galama}, T.~J., {et~al.} 1999,
  \apjl, 523, L33

\bibitem[{{Willingale} {et~al.}(2013){Willingale}, {Starling}, {Beardmore},
  {Tanvir}, \& {O'Brien}}]{Willingale2013a}
{Willingale}, R., {Starling}, R.~L.~C., {Beardmore}, A.~P., {Tanvir}, N.~R., \&
  {O'Brien}, P.~T. 2013, \mnras, 431, 394

\bibitem[{{Wilms} {et~al.}(2000){Wilms}, {Allen}, \& {McCray}}]{Wilms2000}
{Wilms}, J., {Allen}, A., \& {McCray}, R. 2000, \apj, 542, 914

\bibitem[{{Yang} {et~al.}(2022){Yang}, {Ai}, {Zhang}, {Zhang}, {Liu}, {Wang},
  {Yang}, {Yin}, {Li}, \& {L{\"u}}}]{Yang2022}
{Yang}, J., {Ai}, S., {Zhang}, B.-B., {et~al.} 2022, \nat, 612, 232

\bibitem[{{Zhang} {et~al.}(2006){Zhang}, {Fan}, {Dyks}, {Kobayashi},
  {M{\'e}sz{\'a}ros}, {Burrows}, {Nousek}, \& {Gehrels}}]{Zhang2006a}
{Zhang}, B., {Fan}, Y.~Z., {Dyks}, J., {et~al.} 2006, \apj, 642, 354

\end{thebibliography}

\begin{appendix} 
\onecolumn

\section{Afterglow and host galaxy photometry}

\setlength\LTcapwidth{\columnwidth}
\begin{longtable}{cccccc}
\caption{Photometry of the afterglow of GRB\,210610B. (t$_0$ = 19:51:05.05 UT). Magnitudes are given in the AB system and are not corrected for Galactic extinction. We give three magnitudes at 58 days, for the host galaxy measured in an aperture identical to that used in afterglow photometry, for the full host galaxy, and for the companion galaxy, respectively.\label{Tab:phot}} \\
\hline\hline
Epoch           & Band  & Telescope/    & Exposure      & Mag       & Ref   \\ 
t-t$_0$ (day)   &       & Instrument    & (s)           &           &       \\
\hline
\endfirsthead
\caption{Continued.}\\
\hline\hline
Epoch           & Band  & Telescope/    & Exposure      & Mag       & Ref   \\ 
t-t$_0$ (day)   &       & Instrument    & (s)           &           &       \\
\hline
\endhead
\hline
\endfoot
\hline
\endlastfoot
1.09710 & $u^\prime$   & GTC/HiPERCAM      & $10\times60$ & $20.311\pm0.082$ & This Work \\ 
1.12521 & $u^\prime$   & LT/IO:O           & $1\times120$ & $20.430\pm0.082$ & \citep{Perley30216} \\ 
2.08091 & $u^\prime$   & GTC/HiPERCAM      & $10\times60$ & $21.438\pm0.067$ & This Work \\ 
5.18532 & $u^\prime$   & GTC/HiPERCAM      & $10\times60$ & $22.498\pm0.112$ & This Work \\ 
58.1043 & $u^\prime$   & GTC/HiPERCAM      & $30\times60$ & $23.353\pm0.076$ & This Work \\ 
58.1043 & $u^\prime$   & GTC/HiPERCAM      & $30\times60$ & $23.034\pm0.076$ & This Work \\ 
58.1043 & $u^\prime$   & GTC/HiPERCAM      & $30\times60$ & $24.901\pm0.096$ & This Work \\ 
\hline 
0.07316 & $g^\prime$ & Ondrejov D50 & $3\times300$ & $17.597\pm0.018$ & This work \\
0.10523 & $g^\prime$ & Ondrejov D50 & $3\times300$ & $17.627\pm0.019$ & This work \\
0.14793 & $g^\prime$ & Ondrejov D50 & $3\times300$ & $17.652\pm0.023$ & This work \\
0.20578 & $g^\prime$ & Ondrejov D50 & $10\times180$ & $17.833\pm0.083$ &This work \\
0.40536 & $g^\prime$ & 1.22m Palomar P48 Schmidt/ZTF & $1\times30$ & $18.490\pm0.020$ & \cite{Ho2022} \\
0.53296 & $g^\prime$ & 1.22m Palomar P48 Schmidt/ZTF & $1\times30$ & $18.770\pm0.030$ & \cite{Ho2022} \\
1.09710  & $g^\prime$ & GTC/HiPERCAM     & $10\times60$  & $20.010\pm0.040$ & This work \\ 
1.12103 & $g^\prime$ & LT/IO:O     	     & $1\times90$ & $20.040\pm0.050$ & \citep{Perley30216} \\
1.12951 & $g^\prime$ & Ondrejov D50 & $86\times180$ & $20.263\pm0.181$ & This work \\
1.50546 & $g^\prime$ & 1.22m Palomar P48 Schmidt/ZTF & $1\times30$ & $20.750\pm0.140$ & \cite{Ho2022} \\
1.50646 & $g^\prime$ & 1.22m Palomar P48 Schmidt/ZTF & $1\times30$ & $20.870\pm0.120$ & \cite{Ho2022} \\
1.53416 & $g^\prime$ & 1.22m Palomar P48 Schmidt/ZTF & $1\times30$ & $20.800\pm0.140$ & \cite{Ho2022} \\
2.08091 & $g^\prime$ & GTC/HiPERCAM      & $10\times60$ & $21.124\pm0.031$ & This work \\ 
3.15081 & $g^\prime$ & Ondrejov D50 & $97\times180$ & $21.689\pm0.177$ & This work \\
4.13012 & $g^\prime$ & Ondrejov D50 & $78\times180$ & $22.067\pm0.098$ & This work \\
5.18532 & $g^\prime$ & GTC/HiPERCAM      & $10\times60$ & $22.394\pm0.058$ & This work \\ 
6.09497 & $g^\prime$ & Perek 2.0m 	     & $9\times300$ & $22.672\pm0.098$ & This work \\
58.1043 & $g^\prime$ & GTC/HiPERCAM      & $30\times60$ & $23.307\pm0.043$ & This work \\ 
58.1043 & $g^\prime$ & GTC/HiPERCAM      & $30\times60$ & $22.960\pm0.044$ & This work \\ 
58.1043 & $g^\prime$ & GTC/HiPERCAM      & $30\times60$ & $24.866\pm0.053$ & This work \\ 
\hline 
0.04880 & $R$        & FRAM-ORM     & $23\times60$ & $17.211\pm0.065$ & This work \\ 
0.05869 & $r^\prime$ & Ondrejov D50 & $1\times300$ & $17.322\pm0.017$ & This work \\
0.06069 & $R$        & FRAM-ORM     & $8\times60$ & $17.314\pm0.080$ & This work \\ 
0.06219 & $r^\prime$ & Ondrejov D50 & $1\times300$ & $17.377\pm0.014$ & This work \\
0.06568 & $r^\prime$ & Ondrejov D50 & $1\times300$ & $17.441\pm0.019$ & This work \\
0.06642 & $R$        & FRAM-ORM     & $7\times60$ & $17.487\pm0.087$ & This work \\ 
0.06981 & $r^\prime$ & Ondrejov SBT & $17\times120$ & $17.442\pm0.063$ & This work \\
0.07042 & $r^\prime$ & NOT          & $1\times10$ & $17.299\pm0.100$ & \citep{Fynbo2021} \\
0.07176 & $R$        & FRAM-ORM     & $7\times60$ & $17.569\pm0.087$ & This work \\
0.07673 & $R$        & FRAM-ORM     & $6\times60$ & $17.416\pm0.079$ & This work \\
0.08170 & $R$        & FRAM-ORM     & $7\times60$ & $17.489\pm0.078$ & This work \\
0.08290 & $r^\prime$ & Ondrejov SBT & $18\times120$ & $17.489\pm0.066$ & This work \\
0.08667 & $R$        & FRAM-ORM     & $6\times60$ & $17.572\pm0.096$ & This work \\
0.09125 & $R$        & FRAM-ORM     & $6\times60$ & $17.664\pm0.097$ & This work \\
0.09621 & $R$        & FRAM-ORM     & $7\times60$ & $17.534\pm0.084$ & This work \\
0.10119 & $R$        & FRAM-ORM     & $6\times60$ & $17.533\pm0.095$ & This work \\
0.10362 & $r^\prime$ & Ondrejov SBT & $23\times120$ & $17.491\pm0.063$ & This work \\
0.12734 & $r^\prime$ & Ondrejov SBT & $25\times120$ & $17.402\pm0.058$ & This work \\
0.15077 & $r^\prime$ & Ondrejov SBT & $34\times120$ & $17.363\pm0.071$ & This work \\
0.18566 & $r^\prime$ & Ondrejov D50 & $10\times180$ & $17.462\pm0.031$ & This work \\
0.23746 & $r^\prime$ & GTC/OSIRIS   & $1\times30$  & $17.621\pm0.040$ & This work \\
0.44806 & $r^\prime$ & Palomar P48 Schmidt/ZTF & $1\times30$ & $18.229\pm0.020$ & \citep{Ho2022} \\
0.49346 & $r^\prime$ & Palomar P48 Schmidt/ZTF & $1\times30$ & $18.369\pm0.020$ & \citep{Ho2022} \\
1.02539 & $r^\prime$ & CAHA/CAFOS   & $1\times60$   & $19.672\pm0.112$ & This work \\ 
1.03206 & $r^\prime$ & CAHA/CAFOS   & $1\times180$  & $19.653\pm0.035$ & This work \\ 
1.03495 & $r^\prime$ & CAHA/CAFOS   & $1\times180$  & $19.647\pm0.030$ & This work \\ 
1.03791 & $r^\prime$ & CAHA/CAFOS   & $1\times180$  & $19.663\pm0.027$ & This work \\ 
1.09710 & $r^\prime$ & GTC/HiPERCAM & $10\times60$  & $19.720\pm0.050$ & This work \\ 
1.12244 & $r^\prime$ & LT/IO:O      & $1\times90$	& $19.809\pm0.020$ & \citep{Perley30216} \\
1.44226 & $r^\prime$ & Palomar P48 Schmidt/ZTF & $1\times30$ & $20.299\pm0.100$ & \citep{Ho2022} \\
1.56876 & $r^\prime$ & Palomar P48 Schmidt/ZTF & $1\times30$ & $20.239\pm0.100$ & \citep{Ho2022} \\
2.08091 & $r^\prime$ & GTC/HiPERCAM & $10\times60$  & $20.873\pm0.018$  & This work \\ 
2.42766 & $r^\prime$ & Palomar P48 Schmidt/ZTF & $1\times30$ & $21.689\pm0.330$ & \citep{Ho2022} 	\\
2.49606 & $r^\prime$ & Palomar P48 Schmidt/ZTF & $1\times30$ & $21.139\pm0.200$ & \cite{Ho2022} 	\\
5.18532 & $r^\prime$ & GTC/HiPERCAM    & $10\times60$  & $22.315\pm0.042$ & This work \\ 
58.1043 & $r^\prime$ & GTC/HiPERCAM    & $30\times60$  & $23.212\pm0.042$ & This work \\ 
58.1043 & $r^\prime$ & GTC/HiPERCAM    & $30\times60$  & $22.861\pm0.042$ & This work \\ 
58.1043 & $r^\prime$ & GTC/HiPERCAM    & $30\times60$  & $24.486\pm0.058$ & This work \\ 
\hline 
0.09453 & $i^\prime$   & Ondrejov D50      & $5\times180$  & $17.328\pm0.016$ & This work \\
0.16882 & $i^\prime$   & Ondrejov D50      & $5\times180$  & $17.330\pm0.025$ & This work \\
1.09710 & $i^\prime$   & GTC/HiPERCAM      & $10\times60$  & $19.585\pm0.048$ & This work \\ 
1.12383 & $i^\prime$   & LT/IO:O           & $1\times90$  & $19.609\pm0.030$ & \citep{Perley30216} \\
2.08091 & $i^\prime$   & GTC/HiPERCAM      & $10\times60$  & $20.715\pm0.022$ & This work \\ 
5.18532 & $i^\prime$   & GTC/HiPERCAM      & $10\times60$  & $22.145\pm0.047$ & This work \\ 
6.13124 & $i^\prime$   & Ondrejov D50      & $96\times180$  & $>\ 22.760$ & This work \\
58.1043 & $i^\prime$   & GTC/HiPERCAM      & $30\times60$  & $23.052\pm0.023$ & This work \\ 
58.1043 & $i^\prime$   & GTC/HiPERCAM      & $30\times60$  & $22.843\pm0.024$ & This work \\ 
58.1043 & $i^\prime$   & GTC/HiPERCAM      & $30\times60$  & $24.142\pm0.049$ & This work \\ 
\hline 
0.08363 & $z^\prime$   & Ondrejov D50 	   & $5\times180$ & $17.234\pm0.029$ & This work \\
0.11569 & $z^\prime$   & Ondrejov D50      & $5\times180$ & $17.246\pm0.028$ & This work \\
0.15862 & $z^\prime$   & Ondrejov D50      & $5\times180$ & $17.285\pm0.034$ & This work \\
1.0971 & $z^\prime$   & GTC/HiPERCAM      & $10\times60$ & $19.413\pm0.040$ & This work \\ 
1.1269 & $z^\prime$    & LT/IO:O           & $1\times90$  & $19.491\pm0.060$ & \citep{Perley30216} \\
1.1354 & $z^\prime$    & Ondrejov D50      & $9\times300$  & $19.624\pm0.140$ & This work \\
2.0809 & $z^\prime$    & GTC/HiPERCAM      & $10\times60$  & $20.530\pm0.027$ & This work \\ 
5.1853 & $z^\prime$    & GTC/HiPERCAM      & $10\times60$  & $21.953\pm0.052$ & This work \\ 
58.1043 & $z^\prime$   & GTC/HiPERCAM      & $30\times60$  & $23.037\pm0.028$ & This work \\ 
58.1043 & $z^\prime$   & GTC/HiPERCAM      & $30\times60$  & $22.769\pm0.031$ & This work \\ 
58.1043 & $z^\prime$   & GTC/HiPERCAM      & $30\times60$  & $23.620\pm0.057$ & This work \\ 
\hline
0.01173 & Clear & Ondrejov SBT &  $20\times 12$ & $15.931\pm0.218$ & This work \\
0.01504 & Clear & Ondrejov SBT &  $20\times 12$ & $15.451\pm0.128$ & This work \\
0.01809 & Clear & Ondrejov SBT &  $20\times 12$ & $15.743\pm0.141$ & This work \\
0.02108 & Clear & Ondrejov SBT &  $20\times 12$ & $16.243\pm0.136$ & This work \\
0.02399 & Clear & Ondrejov SBT &  $20\times 12$ & $16.419\pm0.120$ & This work \\
0.02705 & Clear & Ondrejov SBT &  $20\times 12$ & $16.640\pm0.128$ & This work \\
0.02988 & Clear & Ondrejov SBT &  $20\times 12$ & $16.685\pm0.099$ & This work \\
0.03266 & Clear & Ondrejov SBT &  $20\times 12$ & $16.860\pm0.113$ & This work \\
0.03550 & Clear & Ondrejov SBT &  $20\times 12$ & $17.047\pm0.119$ & This work \\
0.03835 & Clear & Ondrejov SBT &  $20\times 12$ & $16.954\pm0.095$ & This work \\
0.04113 & Clear & Ondrejov SBT &  $20\times 12$ & $17.075\pm0.099$ & This work \\
0.04391 & Clear & Ondrejov SBT &  $20\times 12$ & $17.002\pm0.088$ & This work \\
0.04668 & Clear & Ondrejov SBT &  $20\times 12$ & $17.083\pm0.096$ & This work \\
0.04953 & Clear & Ondrejov SBT &  $20\times 12$ & $17.187\pm0.096$ & This work \\
0.05238 & Clear & Ondrejov SBT &  $20\times 12$ & $16.975\pm0.075$ & This work \\
0.05886 & Clear & Ondrejov SBT &  $43\times 12$ & $17.233\pm0.059$ & This work \\
\hline
252.3749 & $H$  & GTC/EMIR     &   $349\time3$  & $> 22.9$         & This work \\
\end{longtable}

\pagebreak

\section{Linear polarisation measurements on GRBs afterglow emission.}

\setlength\LTcapwidth{\columnwidth}
\begin{longtable}{cccccc}
\caption{Measured values for the linear polarisation and PA on GRB afterglow from literature.\label{table:plin_theta_values_comparison}} \\
\hline\hline
GRB & Redshift & T$_{mid}$  & P$_{Lin}$ & $\theta$ & Ref\\
    &          & (days)     & ($\%$)    & (º)      &    \\
\hline
\endfirsthead
\caption{Continued.}\\
\hline\hline
GRB & Redshift & T$_{mid}$  & P$_{Lin}$ & $\theta$ & Ref\\
    &          & (days)     & ($\%$)    & (º)      &    \\
\hline
\endhead
\hline
\endfoot
\hline
\hline
\endlastfoot
GRB\,990510  & 1.62 & 0.7708 & 1.7 $\pm$ 0.2 & 101 $\pm$ 3 & \citep{Covino1999} \\
             &      & 0.8583 & 1.6 $\pm$ 0.2 &  96 $\pm$ 4 & \citep{Wijers1999} \\
             &      & 1.8083 & 2.2$_{-0.9}^{+1.1}$ &  112$_{-17}^{+15}$ & \textquotedbl \\
\hline 
GRB\,990712  & 0.43 & 0.44 & 2.9 $\pm$ 0.4 & 121.1 $\pm$  3.5 & \citep{Rol2000} \\
             &      & 0.70 & 1.2 $\pm$ 0.4 & 116.2 $\pm$ 10.1 & \textquotedbl \\
             &      & 1.45 & 2.2 $\pm$ 0.7 & 139.1 $\pm$ 10.4 & \textquotedbl \\
\hline
GRB\,020405  & 0.695 & 1.2292 & 1.50 $\pm$ 0.40 & 172 $\pm$ 8 & \citep{Masetti2003} \\
             &       & 1.3208 & 9.89 $\pm$ 1.30 & 180 $\pm$ 4 & \citep{Bersier2003} \\
             &       & 2.2682 & 1.96 $\pm$ 0.33 & 154 $\pm$ 5 & \citep{Covino2003} \\
             &       & 3.8792 & 1.47 $\pm$ 0.43 & 168 $\pm$ 9 & \textquotedbl \\
\hline
GRB\,020813  & 1.35  & 0.21528  & 2.22 $\pm$ 0.07 & 157.6 $\pm$ 1.0 & \citep{Barth2003}$^{*}$ \\
             &       & 0.26181  & 1.98 $\pm$ 0.04 & 153.4 $\pm$ 1.7 & \textquotedbl \\
             &       & 0.34167  & 1.96 $\pm$ 0.07 & 152.0 $\pm$ 1.2 & \textquotedbl \\
             &       & 0.89792  & 1.07 $\pm$ 0.22 & 154.3 $\pm$ 5.9 & \citep{Gorosabel2004} \\
             &       & 0.93750  & 1.42 $\pm$ 0.25 & 137.0 $\pm$ 4.4 & \textquotedbl \\
             &       & 0.97542  & 1.11 $\pm$ 0.22 & 150.5 $\pm$ 5.5 & \textquotedbl \\
             &       & 1.01625  & 1.05 $\pm$ 0.23 & 146.4 $\pm$ 6.2 & \textquotedbl \\
             &       & 1.11667  & 1.43 $\pm$ 0.44 & 155.8 $\pm$ 8.5 & \textquotedbl \\
             &       & 1.97958  & 1.26 $\pm$ 0.34 & 164.7 $\pm$ 7.4 & \textquotedbl \\            
\hline 
GRB\,021004  & 2.33 & 0.37 & 1.72 $\pm$ 0.56 & 187.7 $\pm$ 8.3 & \citep{Rol2003} \\
             &      & 0.38 & 2.09 $\pm$ 0.60 & 173.0 $\pm$ 7.9 & \textquotedbl \\
\hline
GRB\,030329 & 0.17 & 0.5321 & 0.92 $\pm$ 0.10 & 86.13 $\pm$ 2.43 & \citep{covinogotz16}$^{**}$ \\
            &      & 0.5492 & 0.86 $\pm$ 0.09 & 86.74 $\pm$ 2.40 & \textquotedbl \\
            &      & 0.5671 & 0.87 $\pm$ 0.09 & 88.60 $\pm$ 2.64 & \textquotedbl \\
            &      & 0.5850 & 0.80 $\pm$ 0.09 & 91.12 $\pm$ 2.88 & \textquotedbl \\
            &      & 0.6921 & 0.66 $\pm$ 0.07 & 78.52 $\pm$ 2.94 & \textquotedbl \\
            &      & 0.7129 & 0.66 $\pm$ 0.07 & 76.69 $\pm$ 2.89 & \textquotedbl \\
            &      & 0.7342 & 0.56 $\pm$ 0.05 & 74.37 $\pm$ 3.11 & \textquotedbl \\
            &      & 1.5204 & 1.97 $\pm$ 0.48 & 83.20 & \textquotedbl \\
            &      & 1.5500 & 1.37 $\pm$ 0.11 & 61.65 $\pm$ 2.38 & \textquotedbl \\
            &      & 1.5800 & 1.50 $\pm$ 0.12 & 62.29 $\pm$ 2.44 & \textquotedbl \\
            &      & 1.6700 & 1.07 $\pm$ 0.09 & 59.41 $\pm$ 2.51 & \textquotedbl \\
            &      & 1.7000 & 1.09 $\pm$ 0.08 & 66.07 $\pm$ 2.45 & \textquotedbl \\
            &      & 1.7200 & 1.02 $\pm$ 0.08 & 67.05 $\pm$ 2.60 & \textquotedbl \\
            &      & 1.7400 & 1.13 $\pm$ 0.08 & 70.56 $\pm$ 2.51 & \textquotedbl \\
            &      & 2.6800 & 0.52 $\pm$ 0.06 & 30.76 $\pm$ 5.04 & \textquotedbl \\
            &      & 2.7000 & 0.52 $\pm$ 0.12 & 12.55 $\pm$ 4.63 & \textquotedbl \\
            &      & 2.7200 & 0.31 $\pm$ 0.07 & 24.50 $\pm$ 6.94 & \textquotedbl \\
            &      & 3.5400 & 0.57 $\pm$ 0.09 & 53.85 $\pm$ 4.08 & \textquotedbl \\
            &      & 3.5600 & 0.53 $\pm$ 0.08 & 57.08 $\pm$ 4.06 & \textquotedbl \\
            &      & 3.5800 & 0.42 $\pm$ 0.10 & 62.21 $\pm$ 6.10 & \textquotedbl \\
            &      & 5.6600 & 1.68 $\pm$ 0.18 & 66.32 $\pm$ 3.38 & \textquotedbl \\
            &      & 7.6400 & 2.22 $\pm$ 0.28 & 75.16 $\pm$ 3.32 & \textquotedbl \\
            &      & 9.5900 & 1.33 $\pm$ 0.14 & 70.91 $\pm$ 3.31 & \textquotedbl \\
            &      & 13.6000 & 2.04 $\pm$ 0.57 & 1.16 $\pm$ 7.64 & \textquotedbl \\
            &      & 22.5000 & 0.58 $\pm$ 0.10 & 42.7 $\pm$ 9.26 & \textquotedbl \\
            &      & 37.5000 & 1.48 $\pm$ 0.48 & 25.42 $\pm$ 9.41 & \textquotedbl \\
\hline
XR\,080109   & 0.007 & 3.6416  & 0.95 $\pm$ 0.20 & 114.9 $\pm$ 5.9 & \citep{Gorosabel2010} \\
             &       & 5.5552  & 0.85 $\pm$ 0.28 & 106.1 $\pm$ 9.4 & \textquotedbl \\
             &       & 20.6279 & 1.05 $\pm$ 0.06 & 135.3 $\pm$ 1.7 & \textquotedbl \\
             &       & 20.6443 & 1.28 $\pm$ 0.06 & 132.5 $\pm$ 1.4 & \textquotedbl \\
             &       & 52.5578 & 1.42 $\pm$ 0.46 & 139.0 $\pm$ 9.1 & \textquotedbl \\
\hline
GRB\,080928  & 1.6919 & 1.7  & 4.49$^{+1.16}_{-0.96}$ & 41.3 $\pm$ 6.3 & \citep{Brivio2022} \\
\hline
GRB\,090102  & 1.55  & 0.0025 & 10.1 $\pm$ 1.3 & -- & \citep{Steele2009} \\
\hline
GRB\,091208B & 1.063 & 0.0042  & 10.4 $\pm$ 2.5 & 92 $\pm$ 6 & \citep{Uehara2012} \\
\hline
GRB\,091018  & 0.971 & 0.2461 & 1.07 $\pm$ 0.30 & 179.2 $\pm$ 16.1 & \citep{Wiersema2012} \\
             &       & 0.4548 & 1.44 $\pm$ 0.32 & 2.2 $\pm$ 12.6 & \textquotedbl \\
             &       & 1.1394 & 1.73 $\pm$ 0.36 & 69.8 $\pm$ 11.7 & \textquotedbl \\
             &       & 1.1552 & 3.25 $\pm$ 0.35 & 57.6 $\pm$ 6.1 & \textquotedbl \\
             &       & 1.1735 & 1.99 $\pm$ 0.35 & 27.6 $\pm$ 10.0 & \textquotedbl \\
             &       & 1.1893 & 1.42 $\pm$ 0.36 & 114.6 $\pm$ 14.0 & \textquotedbl \\
             &       & 1.3918 & 0.97 $\pm$ 0.32 & 32.8 $\pm$ 17.8 & \textquotedbl \\
             &       & 1.4493 & 1.08 $\pm$ 0.35 & 88.7 $\pm$ 17.9 & \textquotedbl \\
             &       & 2.3902 & 1.45 $\pm$ 0.37 & 169.0 $\pm$ 14.3 & \textquotedbl \\
\hline
GRB\,120308  & 2.22 & 0.0033 & 28$^{+4}_{-4}$ & 34 $\pm$ 4 & \citep{Mundell2013} \\
             &      & 0.0042 & 23$^{+4}_{-4}$ & 44 $\pm$ 6 & \textquotedbl \\
             &      & 0.0052 & 17$^{+5}_{-4}$ & 51 $\pm$ 9 & \textquotedbl \\
             &      & 0.0062 & 16$^{+7}_{-4}$ & 40 $\pm$ 10 & \textquotedbl \\
             &      & 0.0081 & 16$^{+5}_{-4}$ & 55 $\pm$ 9 & \textquotedbl \\
\hline
GRB\,121024A & 2.298 & 0.2194 & 4.09 $\pm$ 0.2 & 163.7 $\pm$ 2.8 & \citep{Wiersema2014} \\
             &       & 0.2302 & 4.83 $\pm$ 0.2 & 160.3 $\pm$ 2.3 & \textquotedbl \\
             &       & 0.2782 & 3.82 $\pm$ 0.2 & 182.7 $\pm$ 3.0 & \textquotedbl \\
             &       & 0.2928 & 3.12 $\pm$ 0.19 & 175.3 $\pm$ 3.5 & \textquotedbl \\
             &       & 0.3088 & 3.39 $\pm$ 0.18 & 178.0 $\pm$ 2.9 & \textquotedbl \\
             &       & 0.3252 & 3.49 $\pm$ 0.18 & 180.3 $\pm$ 3.0 & \textquotedbl \\
             &       & 0.3412 & 3.2 $\pm$ 0.18 & 174.5 $\pm$ 3.3 & \textquotedbl \\
             &       & 1.2995 & 2.66 $\pm$ 0.6 & 83.0 $\pm$ 12.6 & \textquotedbl \\
\hline
GRB\,191221B & 1.148 & 0.121 & 1.4 $\pm$ 0.1 & 68 $\pm$ 5 & \citep{Urata2023} \\
             &       & 0.417 & 1.0 $\pm$ 0.1 & 57 $\pm$ 5 & \textquotedbl \\
             &       & 2.525 & 1.3 $\pm$ 0.1 & 62 $\pm$ 6 & \textquotedbl \\
\hline
GRB\,210610B & 1.1345 & 0.0973 & 4.27 $\pm$ 1.45 & 183 $\pm$ 9 & This work \\
             &        & 0.2407 & 0.22 $\pm$ 0.20 & 267 $\pm$ 19 & \textquotedbl \\
             &        & 0.2688 & 0.03 $\pm$ 0.17 & -- & \textquotedbl \\
             &        & 0.2793 & 0.15 $\pm$ 0.28 & -- & \textquotedbl \\
             &        & 1.2655 & 2.27 $\pm$ 0.22 & 237 $\pm$ 3 & \textquotedbl \\
             &        & 1.2829 & 1.69 $\pm$ 0.27 & 238 $\pm$ 5 & \textquotedbl \\
\hline
GRB\,210619B & 1.937 & 0.1057 & 2.2 $\pm$ 0.7 & 22.0 $\pm$ 10.0 & \citep{Mandarakas2023} \\
             &       & 0.1070 & 2.6 $\pm$ 0.8 &  2.0 $\pm$  8.0 & \textquotedbl \\
\hline
\hline
\end{longtable}
\tablefoot{We include only the values from literature for which we calculate P/$\sigma_{P}$ $>$ 3.0 with T$_{mid}$ in observer frame. \\
($^{*}$) From the spectropolarimetric measurements in \citep{Barth2003}, since we do not consider chromaticity in the polarisation, we show the median value of the measured polarisation on the different wavelength bins, for the three epochs they presents.\\
($^{**}$) For GRB\,030329 we made use of the data presented in \cite{covinogotz16} and, specifically, the results presented in this review taken from \cite{greiner2003} and \cite{magalhaes2003}.}

\end{appendix}

\end{document}